\pdfoutput=1
\documentclass[acmsmall]{acmart}

\AtBeginDocument{%
  }
\setcopyright{acmlicensed}
\copyrightyear{2024}
\acmYear{2024}
\acmDOI{XXXXXXX.XXXXXXX}

\acmJournal{TOSEM}
\acmVolume{1}
\acmNumber{1}
\acmArticle{1}
\acmMonth{8}

\usepackage{mdframed}
\usepackage{graphicx}
\usepackage{subcaption}
\usepackage{multirow}
\usepackage{multicol}
\usepackage{listings}
\usepackage{xcolor}
\usepackage{tcolorbox}
\usepackage{caption}
\usepackage{amsmath}
\usepackage{amsfonts}
\usepackage{bbold}
\usepackage{amssymb}
\usepackage{algorithm}
\usepackage{algpseudocode}

\definecolor{cgreen}{RGB}{0, 176, 80}

\makeatletter
\def\amsbb{\use@mathgroup \M@U \symAMSb}
\makeatother

\tcbuselibrary{breakable,skins,listings}

\lstdefinestyle{mystyle}{
    backgroundcolor=\color{gray!10},
    commentstyle=\color{green!50!black},
    keywordstyle=\color{blue},
    stringstyle=\color{red},
    basicstyle=\ttfamily\small,
    breaklines=true,
    breakatwhitespace=false,
    showstringspaces=false,
    numbers=left,
    numberstyle=\tiny\color{gray},
    numbersep=5pt,
}

\tcbuselibrary{breakable,skins,listings}

\lstdefinestyle{mystyle}{
    backgroundcolor=\color{gray!10},
    commentstyle=\color{green!60!black},
    keywordstyle=\color{blue},
    numberstyle=\tiny\color{gray},
    stringstyle=\color{red},
    basicstyle=\ttfamily\fontsize{7}{11}\selectfont,
    breaklines=true,
    breakatwhitespace=false,
    showstringspaces=false,
    numbers=left,
    numbersep=5pt,
    tabsize=2,
    frame=single,
    rulecolor=\color{black!30},
    captionpos=b
}

\begin{document}
\title{Fixing Function-Level Code Generation Errors for Foundation Large Language Models}

\author{Hao Wen}
\email{wenhao\_1981@msn.com}
\affiliation{%
  \institution{Chongqing University}
  \city{Chongqing}
  \country{China}
}

\author{Yueheng Zhu}
\email{zhuyueheng@stu.cqu.edu.cn}
\affiliation{%
  \institution{Chongqing University}
  \city{Chongqing}
  \country{China}
}

\author{Chao Liu}
\authornote{corresponding author: Chao Liu (liu.chao@cqu.edu.cn); the first and second authors contributed equally to this work.}
\email{liu.chao@cqu.edu.cn}
\affiliation{%
  \institution{Chongqing University}
  \city{Chongqing}
  \country{China}
}

\author{Xiaoxue Ren}
\email{xxren@zju.edu.cn}
\affiliation{%
  \institution{Zhejiang University}
  \city{Hangzhou}
  \country{China}
}

\author{Weiwei Du}
\email{wenvivid@sina.com}
\affiliation{%
 \institution{Army Logistics Academy}
 \city{Chongqing}
 \country{China}
 }

\author{Meng Yan }
\email{mengy@cqu.edu.cn}
\affiliation{%
  \institution{Chongqing University}
  \city{Chongqing}
  \country{China}
}

\renewcommand{\shortauthors}{Wen et al.}
\begin{abstract}
  Function-level code generation leverages foundation Large Language Models (LLMs) like CodeLlama to automatically produce source code with expected functionality. It has been widely investigated and applied in intelligent programming assistants, such as GitHub Copilot, to enhance software development productivity and reduce repetitive tasks. Despite advancements in foundation LLMs, the generation involves many errors. Existing studies leverage static analysis tools (e.g., TBar and ReCoder) or add another fixing LLM (i.e., CoR and LDB) to post-process these errors. However, there are still many errors remaining to be solved because their root causes have not been investigated yet, making it challenging to design better fixing tools. 
  
  In this paper, we first conducted an empirical study on the generation errors. Specifically, we reproduced 14 representative open/closed source LLMs on the HumanEval dataset and verified their correctness. We obtained 12,837 code generation errors and conducted an in-depth analysis of their causes, leading to 19 categories of error causes. Our empirical analysis indicated that three of these causes can be directly fixed. Based on the findings, we proposed a fixing method called LlmFix, which is a rule-based static analysis approach and addresses these three types of errors through a three-step process: filtering code for indentation correction, truncating redundant generated code, and importing missing modules. Evaluations of LlmFix are conducted from two perspectives: its performance on error-fixing tasks and its impact on improving function-level code generation tasks. For error fixing performance, we built an evaluation dataset LlmErrorEval based on our collected 12k+ errors. Experimental results show that LlmFix achieves a fix rate of 17.1\% outperforming the best state-of-the-art LDB by 8.9\%. For code generation improvements, evaluations of LlmFix on both the HumanEval and MBPP datasets demonstrate its effectiveness, significantly improving code generation accuracy by an average Pass@1 of 7.5\% across 14 LLMs on HumanEval and MBPP.
\end{abstract}



\maketitle

\section{Introduction}\label{intro}
Code generation refers to automatically converting natural language descriptions of software development specifications into code snippets written in a programming language with correct functionality \cite{manning2003code, poesia2022synchromesh}. Intelligent programming assistants like GitHub Copilot \cite{chen2021evaluating} can help developers reduce coding efforts in writing repetitive code functions and improve their development productivity \cite{xu2020incorporating, guo2020graphcodebert}. Essentially, these assistants are powered by Large Language Models (LLMs), such as Codex \cite{chen2021evaluating} and CodeLlama \cite{roziere2023code}. 
These foundation models are pre-trained on extensive codebases and fine-tuned specifically for code generation tasks. To evaluate their performance, researchers have developed various benchmarks, which can be categorized into function-level (e.g., HumanEval \cite{chen2021evaluating} and MBPP \cite{austin2021programsynthesislargelanguage}), class-level (e.g., ClassEval \cite{du2023classevalmanuallycraftedbenchmarkevaluating}), and repo-level (e.g., Swe-bench \cite{jimenez2024swebenchlanguagemodelsresolve}). Function-level benchmarks focus on the correctness of code functionality, while class-level and repo-level benchmarks evaluate code generation that requires the generated code depending on existing APIs implemented in the project for necessary reuse consideration. In this study, we focus on the correctness of code functionality, namely function-level code generation, which is also widely investigated by researchers \cite{chen2021evaluating, austin2021programsynthesislargelanguage} and applied in intelligent programming assistants \cite{Wong_2023, Corso_2024}.

Despite advancements in foundation LLMs, there are still many errors even in function-level code generation \cite{hong2024evaluatingllmsmathematicalcoding, huang2024affectsstabilitytoollearning}. 
To address these errors, Fan et al. \cite{fan2023automated} investigated the viability of fixing them by leveraging existing static analysis tools, e.g., TBar \cite{liu2019tbar} and ReCoder \cite{zhu2021syntax}, but their experimental results indicated that the fixing capability of these tools is ineffective. Recently, researchers tried to add another fixing LLM to address the errors produced by the foundation LLMs, such as Large Language Model Debugger (LDB) \cite{zhong2024ldb} and Chain of Repair (CoR) \cite{wang2023intervenor}. 
Although these methods show better effectiveness, there are still many errors remaining to be solved \cite{atil2024llmstabilitydetailedanalysis}. This is because existing studies have not systematically investigated the root causes of function-level code generation errors in foundation LLMs, making it difficult to develop more effective error-fixing tools.

In this paper, we first conduct an empirical study to evaluate the performance of existing foundation LLMs and identify errors that can be directly fixed. Specifically, we reproduced 14 representative open/closed-source LLMs (e.g., CodeLlama-Python) and verified the reproduction correctness. The experimental results demonstrate that these 14 LLMs exhibit strong reproducibility in performance. Subsequently, we analyzed the root cause of 12,837 errors produced by these 14 LLMs and classified them into 19 categories, suggesting a variety of code generation errors. Besides, we investigated the repairability of these errors according to the given limited context. We found that three categories can be directly fixed, including missing import, redundant generation, and inconsistent indentation.

Based on the empirical findings, we design LlmFix, a rule-based static analysis method specifically tailored to automatically fix the three repairable error categories. Generally, LlmFix works in three steps: \textit{1) Code Filtering}, searching for inconsistent indentation, and replacing it with uniform tab-based indentation. \textit{2) Code Truncation}, discarding redundant code located after the function completion. \textit{3) Importing Missing Modules}, automatically adding corresponding import statements at the beginning of the code when detecting the use of a module that has not been imported.

We conducted a comprehensive evaluation of LlmFix from the following two perspectives: \textit{1) Error Fixing Performance.} We constructed a dataset, LlmErrorEval, for error-fixing tasks based on the collected 12k+ errors to evaluate the performance of existing fixing methods. Experimental results show that LlmFix achieves a fix rate of 17.1\%, outperforming the best baseline LDB by 8.9\%. The fix rate can be further enhanced to 24.9\% by integrating LlmFix with LDB. The integration brings little extra computation overhead as LlmFix only takes 11.5ms to run. \textit{2) Code Generation Improvements.} We applied LlmFix to 14 LLMs for code generation tasks. Experimental results showed that LlmFix is effective for all LLMs and the generation accuracy can be significantly improved by 7.5\% on average.

In summary, the major contributions of this study are:\vspace{3pt}

\begin{itemize}
    \item Conducting a systematic investigation into the performance of foundation LLMs on function-level code generation tasks, categorizing 14 types of errors and analyzing their distributions and 19 types of root causes.\vspace{3pt}
    
    \item Proposing a rule-based static analysis method, LlmFix \cite{replication2024package}, to automatically fix three identified fixable errors: missing imports, redundant generation, and inconsistent indentation.\vspace{3pt}

    \item Constructing a dataset, LlmErrorEval \cite{replication2024package}, comprising 12,837 data samples to evaluate the effectiveness of error-fixing methods.\vspace{5pt}
\end{itemize}

This paper is structured as follows. Section \ref{related} introduces the related work on code generation. Section \ref{empirical} provides the empirical study on LLM reproduction and error analysis. Section \ref{method} presents our error fixing method LlmFix, which is evaluated in Section \ref{evaluate} and discussed in Section \ref{discuss}. Section \ref{conclude} summarizes this paper.

\section{Related Work}\label{related}

\subsection{Function-Level Code Generation}\label{sec_related_code}

The task of code generation has evolved significantly over time, and researchers have developed many solutions \cite{puschel2005spiral, yin2017syntactic, syriani2018systematic, svyatkovskiy2020intellicode, roziere2023code}. Early methods can quickly generate simple code structures based on predefined templates \cite{danilchenko2012automated} or rule systems \cite{bajwa2006rule}. Syriani et al. \cite{syriani2018systematic} categorized template-based code generation approaches, identifying their prevalent characteristics and trends, emphasizing the technique's significance in generating source code across diverse application areas. These simple methods are intuitive to implement and control. However, they have difficulties in implementing complex code and generating code patterns not explicitly defined in the rules \cite{van2018automated}.

As deep learning (DL) techniques evolved, researchers leveraged approaches such as long short-term memory (LSTM) \cite{memory2010long}, to comprehend code semantics. By learning from numerous code data, e.g., from the open-source community GitHub, these models can capture the structural and semantic patterns of programming languages, and generate more complex code \cite{le2022coderl}. Yin et al. \cite{yin2017syntactic} introduced a new neural architecture that incorporates a grammar model to explicitly integrate the target syntax as prior knowledge. Their experiments demonstrate that this method is an effective strategy for scaling up to the generation of complex programs from descriptions in natural language. Generally, the DL-based models can produce syntactically correct code, but the generated code was commonly not fully aligned with the given task description \cite{yang2023deep}.

In recent years, the emergence of LLMs like ChatGPT \cite{brown2020language} and CodeLlama \cite{roziere2023code} have introduced a new phase in code generation. These models are based on the Transformer architecture \cite{vaswani2017attention} and pre-trained on large-scale code datasets to learn the deep semantics and syntactic rules of programming languages. ChatGPT is a representative of closed-source LLMs, which have been widely used and possess strong code generation capabilities (60.30\% on HumanEval) \cite{gpt4result}. In the realm of open-source large models, Rozière et al. \cite{roziere2023code} introduced the CodeLlama series, based on the LLAMA2 architecture, designed to support long context inputs specifically for code generation tasks. Ivison et al. \cite{ivison2023camels} proposed the CodeTulu-2 series, which is based on the CodeLlama model and fine-tuned on a series of high-quality instruction datasets established by them, named TÜLU-V2-mix, achieving better performance than the CodeLlama model. Luo et al. \cite{luo2023wizardcoder} applied the Evol-Instruct method to the code domain, introducing the WizardCoder series, which enables complex instruction fine-tuning. Gunasekar et al. \cite{gunasekar2023textbooks} utilized ``textbook quality" data from the internet and synthesized textbooks and exercises using ChatGPT, training the Phi series models characterized by a small parameter count (1.3B-2.7B) and high code generation performance (50.6\% on HumanEval).

\subsection{Error Fixing for Foundation LLMs}\label{sec_related_error}

Austin et al. \cite{austin2021programsynthesislargelanguage} indicated that existing foundation LLMs involve many generation errors remaining to be solved. There are two types of methods for fixing generation errors.

The first type is the static analysis method. These methods typically rely on program semantics or syntax for code error fixing. Specifically, Smytzek et al. \cite{smytzek2024fixkit} proposed a recent code error fixing framework called FixKit, which combines five widely used code error fixing methods: \textit{1) GENPROG} \cite{weimer2009automatically}: A genetic programming-based approach that evolves program variants to fix bugs by applying mutation operators to the abstract syntax tree of the program; \textit{2) MUTREPAIR} \cite{debroy2010using}: A mutation-based fixing technique that generates patches by applying mutation operators to the faulty program; \textit{3) KALI} \cite{qi2015analysis}: A fixing technique that focuses on removing or modifying potentially faulty code to generate patches; \textit{4) CARDUMEN} \cite{martinez2018ultra}: An automatic program fixing technique that uses a probabilistic model to generate fix templates based on mined code patterns; \textit{5) AE} \cite{weimer2013leveraging}: An adaptive search-based technique that uses program equivalence relations to guide the patch generation process. This framework makes these methods easier to use and compare in terms of performance. However, the authors did not evaluate the effectiveness of these methods in fixing errors in LLM-generated code, which we have addressed in section \ref{evaluate}. Fan et al. \cite{fan2023automated} summarized the characteristics of automatically generated code and evaluated two fixing methods: TBar \cite{liu2019tbar} and Recoder \cite{zhu2021syntax}. TBar represents a search-based and pattern-based fixing tool, whereas Recoder is a learning-based approach. Their experiments showed that these two methods are ineffective in fixing errors in generated code: they have limited search spaces, are unable to generate multi-edit patches, and lack awareness of program dependencies. However, the authors did not propose new effective static analysis methods, which is precisely the focus of our research.

The second type is the fixing LLM. The basic idea of this type is to build another fixing LLM to post-process the code generation errors produced by the foundation LLMs.
Specifically, 
Zhong et al. \cite{zhong2024ldb} introduced the Large Language Model Debugger (LDB) method, an LLM-based fixing approach that allows LLMs to use runtime execution information to improve generated programs and fix errors.
Wang et al. \cite{wang2023intervenor} introduced CoR, a method designed to emulate the human process of debugging. It works by iteratively gathering feedback from compilers, interpreting the causes of errors and crafting solutions in natural language, and ultimately applying fixes to the program.

\section{Empirical Study}\label{empirical}

\subsection{Research Questions}\label{empirical_rqs}
This empirical study aims to evaluate the reproducibility of existing representative foundation LLMs, identify the root causes of code generation errors produced by these LLMs, and determine which errors are repairable to guide potential fixing methods.

To achieve this goal, we investigate the following RQs:

\textbf{RQ1: Can we reproduce the performance of existing foundation LLMs?}
This RQ examines the reproducibility of results claimed in the literature on LLMs for code generation. By attempting to replicate the conditions and methodologies of previous studies, we aim to assess the validity of reported performance metrics. This RQ is crucial for establishing a baseline of expectations when it comes to the operational capabilities of LLMs.

\textbf{RQ2: Why did these foundation LLMs fail in code generation?}
Investigating the code generation errors of LLMs is essential for limitation understanding and further improvements. By running the generated code with test cases and analyzing the occurred errors, we aim to uncover the underlying factors that contribute to these errors.

\textbf{RQ3: Which errors can be directly fixed?} Improving the code generation performance is the final goal of this study. This RQ aims to analyze the possibility of automatically fixing some errors after the code generation without asking users to update prompts or sending regeneration requests.

\subsection{Dataset and Evaluation Measure}\label{sec_method_dataset}

To conduct the empirical study, we utilize the widely used dataset HumanEval \cite{chen2021evaluating}, meticulously crafted by OpenAI, for assessing various LLMs. The HumanEval dataset is a collection of 164 real-world programming challenges. Each generation task involves an average of 7.7 test cases, ensuring the functional correctness of the generated code. Most of the well-known LLMs such as CodeLlama \cite{roziere2023code}, CodeTulu \cite{ivison2023camels}, and WizardCoder \cite{luo2023wizardcoder} utilized HumanEval for evaluating code generation capabilities. The HumanEval comprises five components: \textit{1) task\_id,} a unique identifier for each task, allowing for easy reference and organization of the code generation tasks. \textit{2) prompt,} a textual statement that provides generation requirements for guiding LLMs. \textit{3) canonical\_solution,} the standard solution for each task. \textit{4) test,} a testing function along with several test cases, designed to meticulously assess whether the generated code meets the expected requirement. \textit{5) entry\_point,} the name of a function, which is intended to be the main entry point of the generated code function. 

For this dataset, an LLM needs to generate the functionally correct code, which can pass all the test cases, according to a given prompt. To measure the accuracy of code generation, Chen et al. \cite{chen2021evaluating} provide the $Pass@k$ metric, the percentage of tasks solved by an LLM. A task is considered solved if any top-$k$ code generated by LLM can pass all the test cases. As Pass@k has high variance, we follow Chen et al. \cite{chen2021evaluating} and use the unbiased version: $Pass@k=\mathbb{E}\left[1-\binom{n-c}{k}/\binom{n}{k}\right]$, where $\mathbb{E}$ denotes the average performance across all tasks; $n$ is the total number of tasks; $c$ is the number of correctly solved tasks; $\binom{n}{k}$ is the number of combinations for choosing $k$ tasks out of $n$; $\binom{n-c}{k}$ is the number of combinations for choosing $k$ tasks from those one model cannot solve (i.e., $n-c$ tasks), representing the number of ways to fail. Note that this study measured the code generation performance with $Pass@1$, as developers commonly expected that LLMs could generate correct code in the first place. Besides, to mitigate the effects of randomness in LLMs, we follow \cite{liu2021reproducibility} to test each LLM ten times and then report the average value as the final $Pass@1$ result.

\subsection{Can We Reproduce the Performance of Existing Foundation LLMs? (RQ1)} \label{empirical_rq1}
\subsubsection{Motivation and Method}
In recent years, researchers have developed many foundation LLMs and published their test results on the HumanEval dataset \cite{zheng2023survey}. However, achieving outcomes that are completely identical to those reported remains a challenging task \cite{chang2024survey}. Liu et al. \cite{Liu_2021} conducted a literature review on DL studies published in twenty prominent SE venues over the past five years. Their findings indicate that DL models inherently possess randomness, which can compromise the replicability of studies involving unstable models. A DL model with a low level of convergence can exhibit highly variable and unstable performance, resulting in poor reproducibility of the studies. Therefore, to thoroughly investigate whether existing LLMs are consistent with the reported performances as described in RQ1, we adopted several LLMs for testing. One is the ChatGPT, a closed-source LLM released by OpenAI \cite{ChatGPT}. 

For open-source LLMs, Zheng et al. \cite{zheng2023survey} reported on the performance of many LLMs on HumanEval. Based on this study, we selected four families of LLMs. Phind \cite{phind2023} performs best in the reported ranking, followed by WizardCoder \cite{luo2023wizardcoder} and CodeLlama \cite{roziere2023code}. Phi \cite{gunasekar2023textbooks} is the best small-sized LLM and is competitive with other families. Besides, we included the CodeTulu family \cite{ivison2023camels} in our study due to its promising performance as reported by Ivison et al. \cite{ivison2023camels}. All open-source LLMs are re-implemented by using the LLM parameters stored in the HuggingFace platform and tested on a server with four NVIDIA RTX3090 GPUs in default settings.

Detailed information about each LLM family and relevant links are listed below. We ran each LLM ten times on HumanEval and compared the code generation performance from original reports \cite{zheng2023survey,ivison2023camels}.

\begin{itemize} 
    \item[$\bullet$] \textbf{ChatGPT,} is a powerful LLM released by OpenAI \cite{openai2024gpt4technicalreport}. Researchers have demonstrated its high performance in code generation \cite{10113601}. We chose the GPT-3.5-turbo version with 175B parameters by invoking official APIs \cite{APIKey} following Sun et al. \cite{sun2023does}. 

    \item[$\bullet$] \textbf{Phind,} was introduced by the Phind team, available in two versions \cite{huggingfacephind}: Phind-CodeLlama-34B-v1 and Phind-CodeLlama-34B-v2. According to the team's report \cite{phind2023}, the latter version even outperforms OpenAI's GPT series in code generation. 

    \item[$\bullet$] \textbf{WizardCoder}, was introduced by Microsoft \cite{luo2023wizardcoder}, which empowers code generation LLMs with complex instruction fine-tuning. We re-implemented the versions of WizardCoder-Python in different parameter sizes (7B/13B/34B) \cite{huggingfacewizardcoder}. 
     
    \item[$\bullet$] \textbf{CodeLlama}, was introduced by Meta AI \cite{roziere2023code}, which is a competitive code generation model and inspired many other varied versions such as CodeTulu \cite{ivison2023camels} and WizardCoder \cite{luo2023wizardcoder}. We used the CodeLlama-python (7B/13B/34B) \cite{huggingfacecodellama} as our baseline, which fine-tunes CodeLlama with more Python code and shows better performance in code generation.

    \item[$\bullet$] \textbf{Phi,} was released by Microsoft \cite{gunasekar2023textbooks}, which is renowned for its smaller model size (1.3B) but demonstrated exceptional high performance in code generation. Later, Microsoft released new versions including Phi-1.5  Li et al. \cite{li2023textbooks} and Phi-2 \cite{phi-2-2023}. Thus, these three models were selected as our baselines including Phi-1 (1.3B), Phi-1.5 (1.3B), and Phi-2 (2.7B) \cite{huggingfacephi}.

    \item[$\bullet$] \textbf{CodeTulu}, was proposed by AllenAI \cite{huggingfacecodetulu}, whose coding capabilities matched or even exceeded the performance of GPT-3.5-turbo in several benchmark tests. We selected the best-performing and variably parameterized versions of CodeTulu-2 (7B/13B/34B) as our baselines.
\end{itemize}

\subsubsection{LLMs Reproduction and Evaluation}
\begin{figure}
    \centering
    \includegraphics[width=0.5\linewidth]{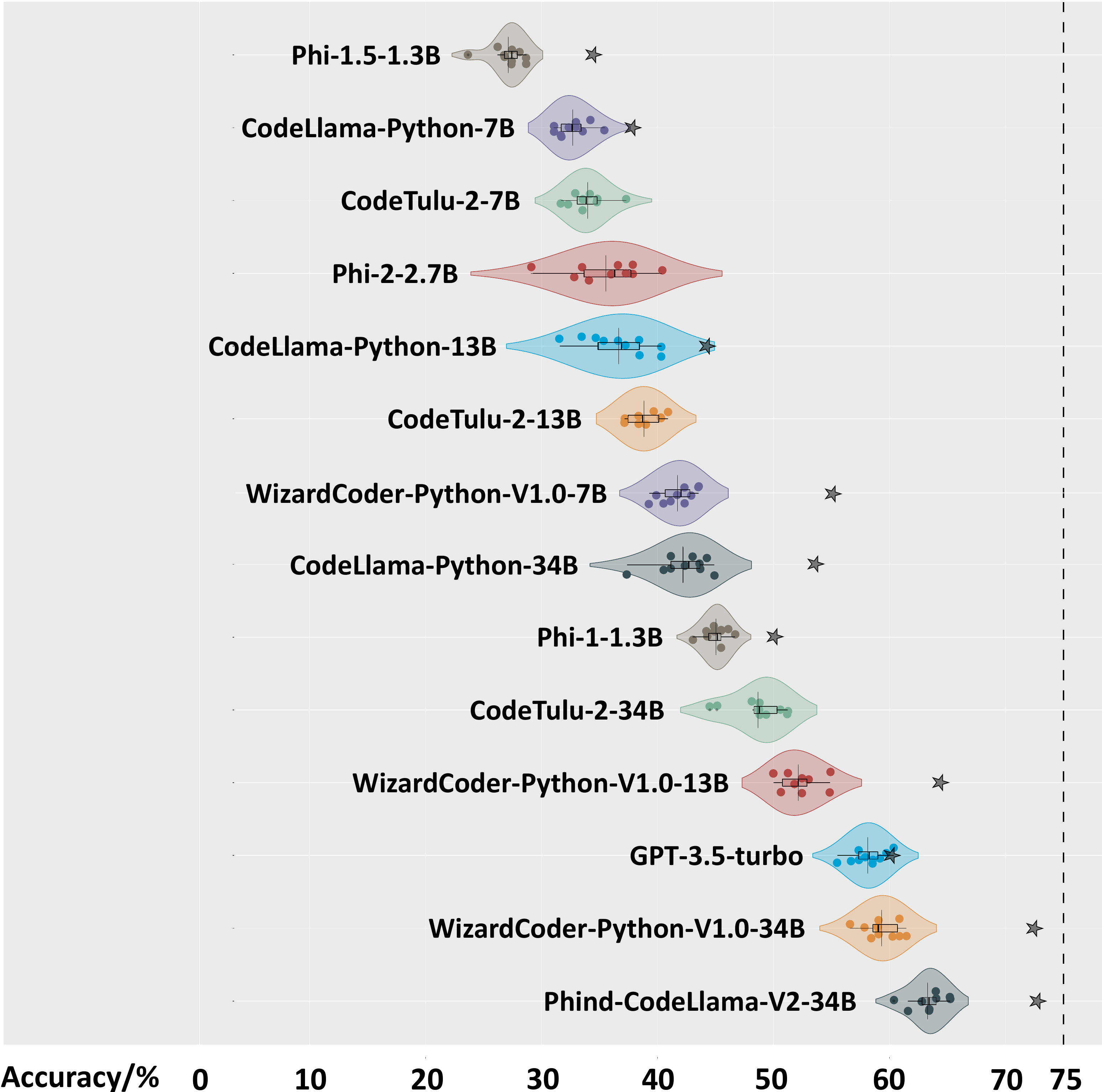}
    \caption{The distribution of test results for 14 models, sorted by the mean values from lowest to highest, where the thin vertical lines represent the means; the thick short lines inside the boxes represent the medians; and the stars indicate the reported performance in the related papers \cite{zheng2023survey}. }
    \label{fig_violin}
\end{figure}

Fig. \ref{fig_violin} shows the accuracy distribution of 14 LLMs over ten times tests, sorted by the average accuracy from lowest to highest. The best-performing model is the Phind-CodeLlama-V2-34B, with an average accuracy of 63.29\%. Table \ref{tab_val} lists the specific numerical characteristics of these 14 LLMs from the ten tests, including minimum, maximum, mean, standard deviation, and results reported in related research papers \cite{zheng2023survey}. From the study by Zheng et al. \cite{zheng2023survey}, we know that the 14 LLM accuracies range from 34.10\% to 73.80\%, while our replicated test results have an average accuracy ranging from 27.14\% to 63.29\%. Compared to reported data, our average test results are approximately reaching 84.07\% of the reported results effectively. This difference is within an acceptable range, likely related to the numerical precision used in the experiments. Additionally, the standard deviation (STD) for all models is significantly lower than the mean (less than 10\% of the mean), indicating low data variability. 

\begin{table*}
    \centering
    \small
    \caption{Accuracy of ten tests of LLMs on HumanEval dataset in terms of minimum, maximum, mean, and standard deviation (STD) of Pass@1 values, compared to the result reported in related papers \cite{zheng2023survey}, where the percentage value in parenthesis indicates the improvement ratio to the reported; ``-" indicates the related paper did not report the Pass@1 as required in our study but Pass@10.}
    \begin{tabular}{lllllc}
        \hline\noalign{\vspace{1pt}}
        \textbf{Model} & \textbf{Reported} & \textbf{Minimum} & \textbf{Maximum} & \textbf{Mean} & \textbf{STD} \\
        \hline\noalign{\vspace{1pt}}
        CodeLlama-Python-7B         & 38.40 & 31.10 (\textcolor{cgreen}{$\downarrow$19.01\%}) & 35.37 (\textcolor{cgreen}{$\downarrow$07.89\%}) & 32.69 (\textcolor{cgreen}{$\downarrow$14.87\%}) & 1.39  \\
        CodeLlama-Python-13B        & 43.30 & 31.71 (\textcolor{cgreen}{$\downarrow$26.77\%}) & 40.24 (\textcolor{cgreen}{$\downarrow$07.07\%}) & 36.65 (\textcolor{cgreen}{$\downarrow$15.36\%}) & 2.82  \\
        CodeLlama-Python-34B        & 53.70 & 39.02 (\textcolor{cgreen}{$\downarrow$27.34\%}) & 46.34 (\textcolor{cgreen}{$\downarrow$13.71\%}) & 43.72 (\textcolor{cgreen}{$\downarrow$18.58\%}) & 2.17  \\
        WizardCoder-Python-V1.0-7B  & 55.50 & 40.05 (\textcolor{cgreen}{$\downarrow$27.75\%}) & 43.90 (\textcolor{cgreen}{$\downarrow$20.90\%}) & 42.23 (\textcolor{cgreen}{$\downarrow$23.91\%}) & 1.05  \\
        WizardCoder-Python-V1.0-13B & 64.00 & 48.17 (\textcolor{cgreen}{$\downarrow$24.73\%}) & 56.10 (\textcolor{cgreen}{$\downarrow$12.34\%}) & 52.13 (\textcolor{cgreen}{$\downarrow$18.54\%}) & 2.48  \\
        WizardCoder-Python-V1.0-34B & 73.20 & 57.32 (\textcolor{cgreen}{$\downarrow$21.72\%}) & 62.80 (\textcolor{cgreen}{$\downarrow$14.21\%}) & 59.09 (\textcolor{cgreen}{$\downarrow$19.29\%}) & 1.58  \\
        Phind-CodeLlama-V2-34B      & 73.80 & 60.37 (\textcolor{cgreen}{$\downarrow$18.20\%}) & 65.24 (\textcolor{cgreen}{$\downarrow$11.60\%}) & 63.29 (\textcolor{cgreen}{$\downarrow$14.24\%}) & 1.51  \\
        Phi-1-1.3B                  & 50.60 & 43.29 (\textcolor{cgreen}{$\downarrow$14.45\%}) & 46.95 (\textcolor{cgreen}{$\downarrow$07.21\%}) & 45.30 (\textcolor{cgreen}{$\downarrow$10.47\%}) & 1.04  \\
        Phi-1.5-1.3B                & 34.10 & 23.78 (\textcolor{cgreen}{$\downarrow$30.26\%}) & 28.66 (\textcolor{cgreen}{$\downarrow$15.95\%}) & 27.14 (\textcolor{cgreen}{$\downarrow$20.41\%}) & 1.42  \\
        Phi-2-2.7B                  & -     & 29.27 & 40.24 & 35.55 & 3.13  \\
        GPT-3.5-turbo               & 60.30 & 55.49 (\textcolor{cgreen}{$\downarrow$07.98\%}) & 60.37 (\textcolor{red}{$\uparrow$00.12\%}) & 58.11 (\textcolor{cgreen}{$\downarrow$03.63\%}) & 1.46  \\
        CodeTulu-2-7B               & -     & 31.71 & 37.20 & 33.97 & 1.55  \\
        CodeTulu-2-13B              & -     & 37.20 & 40.85 & 38.84 & 1.38  \\
        CodeTulu-2-34B              & -     & 44.51 & 51.22 & 48.66 & 2.30  \\
        \hline
    \end{tabular}
    \label{tab_val}
\end{table*}

To assess reproducibility, we calculated the Pearson correlation coefficient \cite{pearson1895vii} between our mean values and the reported values. The correlation coefficient of about $0.989$ indicates almost perfect consistency between our results and the reported data. The corresponding P-value of $6.91 \times 10^{-9}$ is highly significant statistically, demonstrating a very high level of replication. Phi-2 and the CodeTulu-2 family of LLMs did not report the usual Pass@1 values. Instead, the CodeTulu-2 family chose to report Pass@10 to highlight the superiority of their models. In our study, these models had an average accuracy of 39.26\%, ranking ninth among all models. 

\vspace{5pt}
\begin{mdframed}[nobreak=true]
\textbf{Answer to RQ1:} All our reproduced foundation LLMs exhibit close performance as the original reports: compared to reported data, our average test results are reaching 84.1\% of the reported results effectively.
\end{mdframed}

\subsection{Why did These LLMs Fail in Code Generation? (RQ2)}\label{sec_method_analysis} \label{empirical_rq2}

\subsubsection{Motivation and Method}
To investigate why these LLMs failed in code generation, we extracted the error information from failed test cases and developed a systematic process for error analysis. The HumanEval dataset consists of 164 code generation tasks, each accompanied by an average of 7.7 corresponding test cases \cite{chen2021evaluating}. Our objective is to take the generated code and test cases as input and return the test results, assessing whether the code passed or failed. Utilizing the model and dataset interfaces provided by HuggingFace, we obtained the code generated by the model for each sample in the dataset, along with the corresponding test cases, all presented in string format. These strings can be parsed and executed as a series of Python statements using Python's built-in \texttt{exec()} function, thus enabling automated testing. If the generated code passes all test cases, the test continues running; otherwise, the generated code fails one or more test cases, reports an exception, and terminates.

\begin{figure}
    \centering
    \begin{subfigure}[b]{\linewidth}
    \centering
        \includegraphics[width=\linewidth]{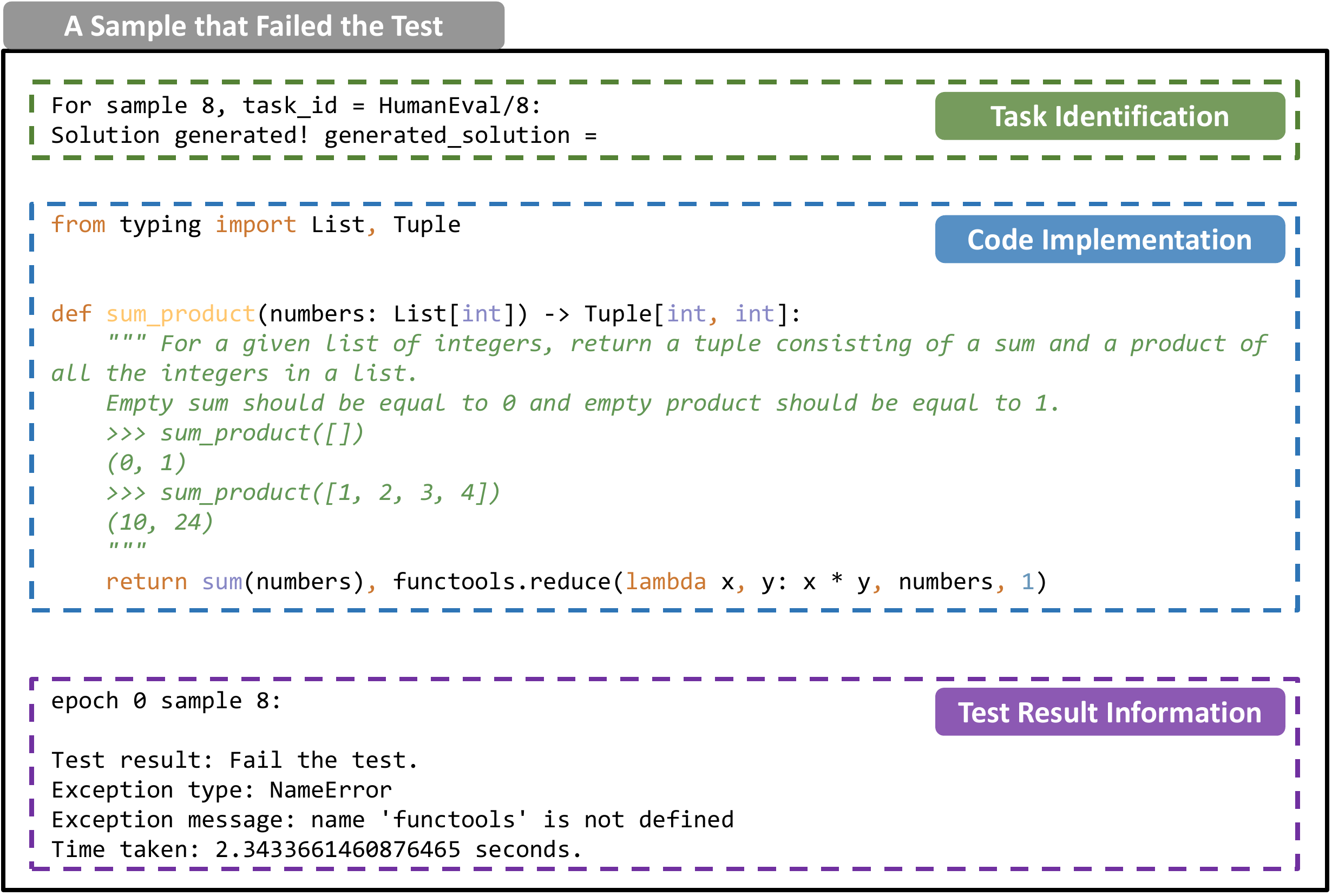}
        \label{fig_fail_sample}
    \end{subfigure}

    \caption{After testing each sample in the dataset, the output file displays the structure of the samples, including an example of a sample that passed the test and another that did not pass the test.}
    \label{fig_test_samples}
\end{figure}

To facilitate batch retrieval of test results and reported exception messages for in-depth error analysis, we employed \texttt{try-except} statements, placing the execution of the generated code and test cases within a \texttt{try} block to catch and record any exceptions. This approach allowed to capture of detailed information about the errors, including the exception type and specific exception information. The extracted information was structured into a specific data format to facilitate further analysis. As shown in Fig. \ref{fig_test_samples}, it consists of three main parts: \textit{1) Task Identification:} the current sample's number and an indication of whether the model successfully generated code. \textit{2) Code Implementation:} the code content generated by the model, based on the prompt part of the dataset, which is a completion of the original prompt. \textit{3) Test Result Information:} a detailed explanation of the test results for the code generated by the model, including whether the test passed or failed. If the test fails, it also provides the error type and specific error information that caused the error.

\subsubsection{Error Classification and Causes Analysis}
Through ten rounds of testing on 14 different LLMs using the HumanEval dataset, we collected a total of 12,837 errors. Based on the \textit{Test Result Information} shown in Fig. \ref{fig_test_samples} which are defined and reported by compiler, we collected these errors for a total of 14 Python exception types. Fig. \ref{fig_Stereo-bar} presents a three-dimensional bar chart illustrating the frequency of occurrence for each of these 14 error types across all LLMs over the ten test rounds. Moreover, considering that the same type of error can be triggered by various causes, we unanimously agreed that it is necessary not only to categorize errors but also to analyze the code containing errors, thereby identifying one or more causes leading to each type of error. The above research results are summarized in Table \ref{tab_error}. The following is a detailed explanation of each error type:

\begin{figure*}
    \centering
    \includegraphics[width=\linewidth]{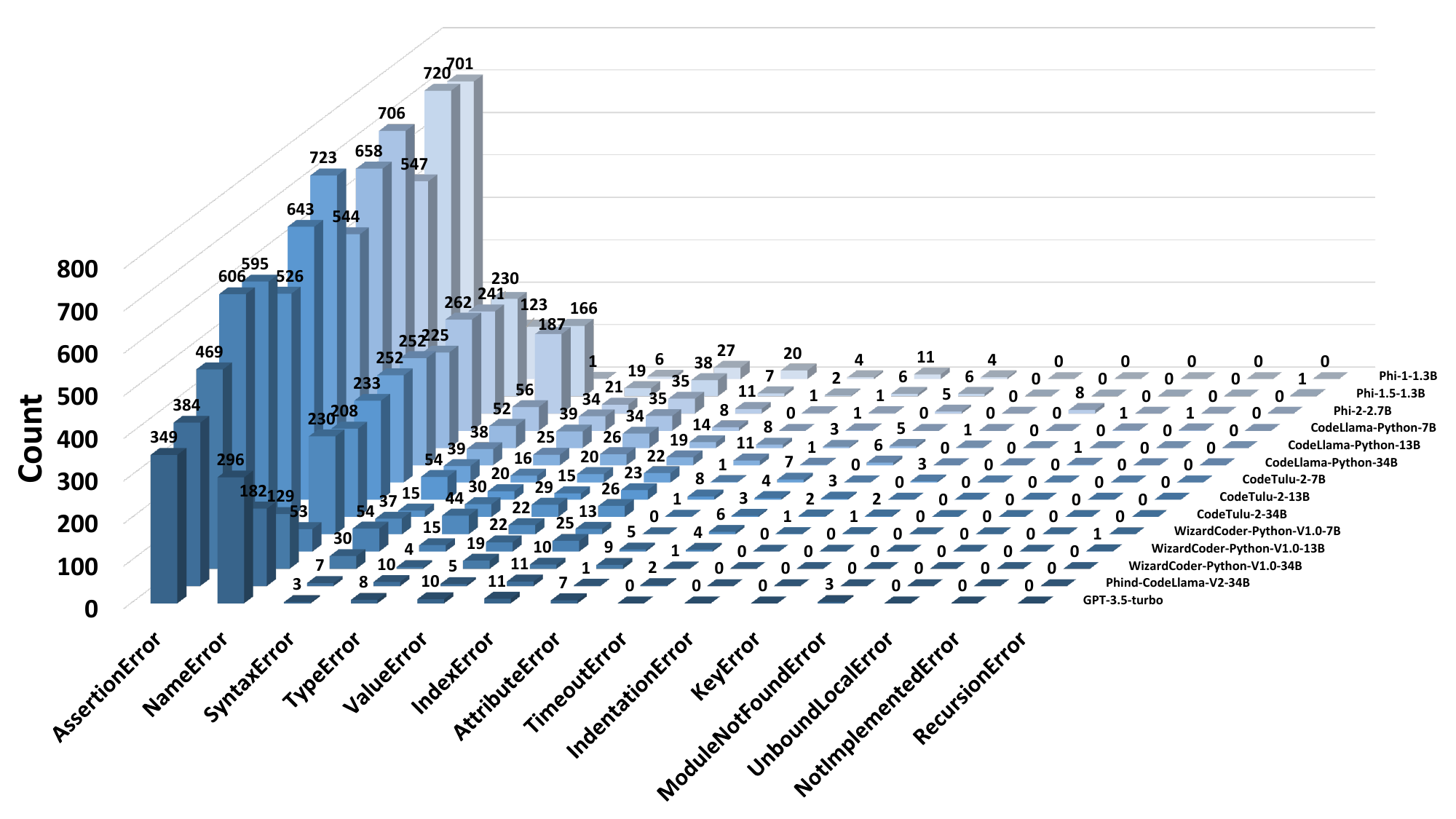}
    \caption{Distribution of total errors across 14 LLMs in ten test runs.}
    \label{fig_Stereo-bar}
\end{figure*}

\begin{table*}
    \footnotesize
    \centering
    \caption{Error types and causes summary: categories, frequencies, causes, and code examples. Note that examples are provided in the Appendix.}
    \label{tab_error}
    \begin{tabular}{p{70pt}p{40pt}p{80pt}p{125pt}r}
        \hline\noalign{\vspace{1pt}}
        \textbf{Error Type} & \textbf{\#Errors} & \textbf{Error Cause} & \textbf{One-line Description} & \textbf{Example} \\
        \hline\noalign{\vspace{1pt}}
        \multirow{4}{*}{AssertionError} & \multirow{4}{*}{8171 (63.64\%)} & Test Case Failure & Model-generated function fails pre-set test cases & listing \ref{code_assertion1}  \\
        & & Empty Function & Model creates an empty function or uses pass statement & listing \ref{code_assertion2}, \ref{code_assertion3}  \\
        \hline\noalign{\vspace{1pt}}
        \multirow{6}{*}{NameError} & \multirow{6}{*}{2916 (22.71\%)} & Misremembered Name & Model defines a function but calls it with incorrect name & listing \ref{code_name1}  \\
        & & Missing Content & Model doesn't generate any content, leading to missing entry\_point & -\footnote{The "Missing content" error occurs when the language model fails to generate any code at all. In such cases, there is no code snippet to display as an example. This error is particularly challenging to illustrate but is included due to its significance in understanding model behavior.}  \\
        & & Missing Import & Model doesn't import required modules or functions & listing \ref{code_name2} \\
        \hline\noalign{\vspace{1pt}}
        \multirow{5}{*}{SyntaxError} & \multirow{5}{*}{739 (5.76\%)} & Unbalanced Delimiters & Imbalance in quotes or parentheses in generated code & listing \ref{code_syntax1}  \\
        & & Function Overflow & Excessive function generation hits limit, causing incomplete output and SyntaxError. & listing \ref{code_syntax2} \\
        \hline\noalign{\vspace{1pt}}
        \multirow{6}{*}{ValueError} & \multirow{6}{*}{337 (2.63\%)} & Empty Sequence & Function fails to handle empty input, causing ValueError & listing \ref{code_value1}  \\
        & & Intentional Raise & Model includes raise statements for program robustness & listing \ref{code_value2}  \\
        & & Inappropriate Argument & Function receives correct type but inappropriate value & listing \ref{code_value3}  \\
        \hline\noalign{\vspace{1pt}}
        IndexError & 291 (2.27\%) & Out of Bounds & Attempting to access an index outside sequence range & listing \ref{code_index1}  \\
        \hline\noalign{\vspace{1pt}}
        TypeError & 220 (1.71\%) & Incompatible Operation & Applying operation to object of inappropriate type & listing \ref{code_type1}  \\
        \hline\noalign{\vspace{1pt}}
        AttributeError & 58 (0.45\%) & Non-existent Attribute & Accessing a non-existent attribute of an object & listing \ref{code_attr1}  \\
        \hline\noalign{\vspace{1pt}}
        TimeoutError & 50 (0.39\%) & Execution Timeout & Code execution exceeds set time limit & listing \ref{code_time1}  \\
        \hline\noalign{\vspace{1pt}}
        IndentationError & 32 (0.25\%) & Inconsistent Indentation & Indentation levels in same code block are inconsistent & listing \ref{code_inden1} \\
        \hline\noalign{\vspace{1pt}}
        ModuleNotFoundError & 11 (0.09\%) & Missing Module & Importing a non-standard library module that's not installed & listing \ref{code_module}  \\
        \hline\noalign{\vspace{1pt}}
        KeyError & 7 (0.05\%) & Non-existent Key & Attempting to access a non-existent key in a dictionary & listing \ref{code_key}  \\
        \hline\noalign{\vspace{1pt}}
        UnboundLocalError & 2 (0.02\%)  & Unassigned Variable & Referencing a local variable before assignment & listing \ref{code_Unbound}  \\
        \hline\noalign{\vspace{1pt}}
        RecursionError & 2 (0.02\%)  & Infinite Recursion & Recursive function lacks proper termination condition & listing \ref{code_recur1}  \\
        \hline\noalign{\vspace{1pt}}
        NotImplementedError & 1 (0.01\%)  & Intentional Raise & Model includes raise statements for program robustness & listing \ref{code_not}  \\
        \hline
    \end{tabular}
\end{table*}

\textit{AssertionError.} This error occurs when an \texttt{assert()} statement fails. It can be attributed to two specific causes. The first is ``Test case failure'', which occurs when the model-generated function fails to pass the test cases in the HumanEval dataset (as illustrated in Listing \ref{code_assertion1}). The second is ``Empty function'', which results from the model creating an empty function or using only a \texttt{pass} statement (Listings \ref{code_assertion2} and \ref{code_assertion3}).

\textit{NameError.} This error occurs when a local or global name is not found. It can be attributed to three specific causes. ``Misremembered name'' occurs when the model defines a function but calls it with an incorrect name (Listing \ref{code_name1}). ``Missing content'' happens when the model doesn't generate any content, leading to a missing entry\_point. Due to the nature of this error, there is no code to display. ``Missing import'' is caused by the model not importing required modules or functions (Listing \ref{code_name2}).

\textit{SyntaxError.} This error occurs when the parser encounters a syntax error. It can be attributed to two specific causes. ``Unbalanced delimiters'' occur due to an imbalance in quotes or parentheses in the generated code (Listing \ref{code_syntax1}). ``Function Overflow'' happens when the excessive function generation hits limit, causing incomplete output and \textit{SyntaxError} (Listing \ref{code_syntax2}).

\textit{ValueError.} This error occurs when a function receives an argument of the correct type but an inappropriate value. It can be attributed to three specific causes. ``Empty sequence'' occurs when a function fails to handle empty input (Listing \ref{code_value1}). ``Intentional raise'' happens when the model includes raising statements for program robustness (Listing \ref{code_value2}). ``Inappropriate argument'' is when a function receives a correct type but inappropriate value (Listing \ref{code_value3}).

Due to the lower frequency of occurrence for the following error, we had fewer samples to analyze, resulting in only one specific cause for each. \textit{IndexError} occurs when attempting to access an index outside the sequence range (Listing \ref{code_index1}). \textit{TypeError} happens when applying an operation to an object of an inappropriate type (Listing \ref{code_type1}). \textit{AttributeError} is raised when accessing a non-existent attribute of an object (Listing \ref{code_attr1}). \textit{TimeoutError} occurs when code execution exceeds the set time limit (Listing \ref{code_time1}). \textit{IndentationError} happens when indentation levels in the same code block are inconsistent (Listing \ref{code_inden1}). \textit{ModuleNotFoundError} is raised when importing a non-standard library module that's not installed (Listing \ref{code_module}). \textit{KeyError} occurs when attempting to access a non-existent key in a dictionary (Listing \ref{code_key}). \textit{UnboundLocalError} happens when referencing a local variable before assignment (Listing \ref{code_Unbound}). \textit{RecursionError} is raised when a recursive function lacks a proper termination condition (Listing \ref{code_recur1}). Finally, \textit{NotImplementedError} occurs when the model includes raise statements for program robustness (Listing \ref{code_not}), which is consistent with one of the causes leading to ValueError.

\subsubsection{Distribution of Error Types} 
\begin{figure*}
    \centering
    \includegraphics[width=\textwidth]{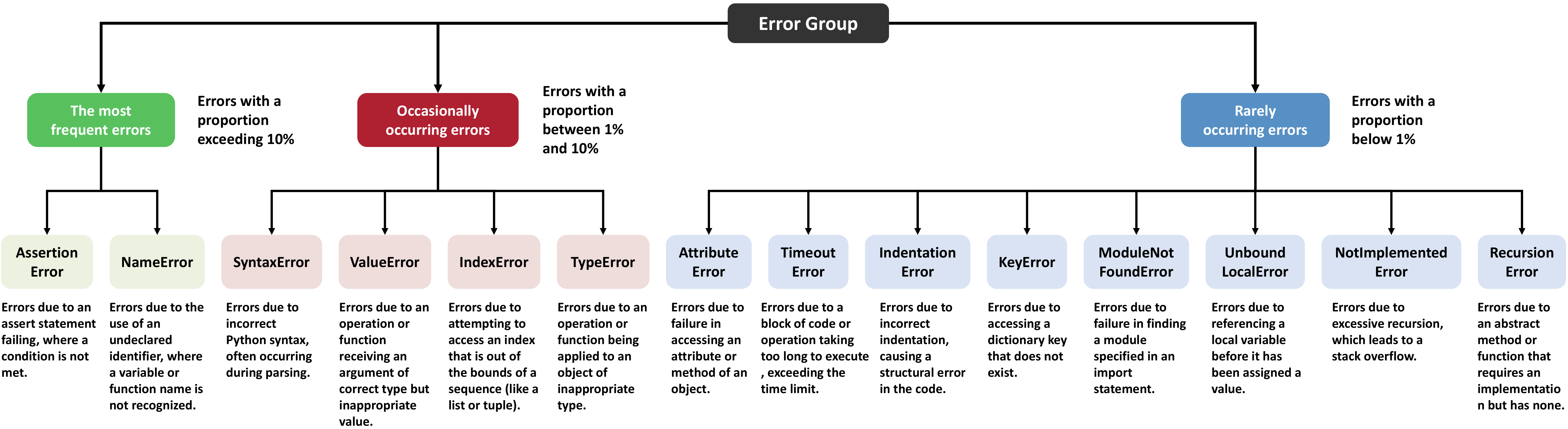}
    \caption{Taxonomy of errors introduced while generating code using LLM.}
    \label{fig_exception_type}
\end{figure*}

Based on Table \ref{tab_error}, we have found that the 14 types of errors can be categorized into three groups based on their frequency of occurrence. The groups and descriptions of the 14 types of errors are illustrated in Fig. \ref{fig_exception_type}.

\subsubsection{The relationship between failed samples and errors} A single piece of code may contain multiple errors. However, the Analysis Process described in \ref{sec_method_analysis} is currently limited to extracting only the first error information from the code. 
To address this limitation, we apply a ``fix-test'' cycle that continually fixes the erroneous code until it passes all the test cases. By applying this cycle to code generated by LLMs, we confirmed that a single code snippet can contain multiple errors, establishing a one-to-many relationship between failed code snippets and errors.
For example, the prompt from the HumanEval dataset (task\_id 133), which serves as the comment in Listing \ref{list_ft1}, specifies the task for the LLM: squaring each number in a given list of digits and rounding each element to the upper integer (ceiling) first. The original code generated by the model, shown in Listing \ref{list_ft1}, triggered a \textit{SyntaxError} due to an unclosed ``[''. Based on the error type and the information provided, we manually corrected the error, resulting in the code shown in Listing \ref{list_ft2}. Subsequent testing revealed that this corrected code still failed to pass all tests. The error occurred because a generator expression was passed into the \texttt{int()} function, which expects a string, bytes, or a real number, leading to a \textit{TypeError}. Consequently, we performed a second manual error correction, resulting in the code shown in Listing \ref{list_ft3}. After this final correction, the code successfully passed all test cases. 

\vspace{5pt}
\begin{mdframed}[nobreak=true]
\textbf{Answer to RQ2: } We collected a total of 12,837 errors, which can be classified into 14 categories according to Python's built-in exception types. Among these, AssertionError emerged as the most common error type. The causes of these errors can be attributed to 19\footnote{The ``Error Cause'' column in Table \ref{tab_error} has 20 rows, but both ValueError and NotImplementError can be caused by the same cause, ``Intentional Raise''. Therefore, we have 19 distinct error causes in total.} different causes, such as incorrect function implementation, syntax mistakes, logical errors, and failure to handle edge cases.
\end{mdframed}

\subsection{Which Errors Can be Directly Fixed? (RQ3)}\label{method_rq2_step3}\label{empirical_rq3}

\subsubsection{Motivation and Method}
This subsection describes the process of manually analyzing errors from LLM-generated code to determine which types are suitable for automated fixing. The approach involves sampling from the errors obtained from the test on the HumanEval dataset, analyzing these samples, and then establishing criteria for automatically fixable errors.

We tested 14 LLMs, running each model ten times on the HumanEval dataset. This process resulted in a total of 12,837 errors. For each error type, we employed a stratified sampling approach to ensure comprehensive representation. For categories with fewer than 100 samples, we included all available examples. For categories with 100 or more samples, we sampled 30\% of the total, with a minimum of 100 and a maximum of 200 samples per category. This adaptive sampling strategy allowed us to maintain a balance between thoroughness and manageability across varied error frequencies.

After analyzing these sampled errors, we engaged in extensive discussions between two authors and solicited a wide range of opinions. We unanimously established the following criteria for errors that can be automatically fixed:

\begin{itemize}
    \item \textbf{Consistency and Predictability.} The error exhibits high consistency and predictability, manifesting similarly across different code instances, and can be addressed using general rules or patterns. This contrasts with errors that are highly dependent on specific code logic and require an understanding of particular code context for resolution.
    
    \item \textbf{Locality.} The error is confined to a specific part of the code and requires only local context information for fixing. This is in contrast to errors that may involve interactions across an entire function or between multiple functions, necessitating broader context or external knowledge, such as understanding the overall problem logic.
    
    \item \textbf{Complexity of Fix.} The error can be resolved through simple rules or pattern matching. This is opposed to errors requiring complex reasoning or multi-step modifications for correction.
\end{itemize}

\begin{table*}
    \centering
    \caption{Analysis of error causes for automatic fixability based on criteria mentioned in subsection \ref{method_rq2_step3}}
    \small
    \label{tab_fixable}
    \begin{tabular}{lcccc}
        \hline\noalign{\vspace{1pt}}
        \textbf{Error Cause} & \textbf{Consistency and Predictability} & \textbf{Locality} & \textbf{Low Complexity} & \textbf{Fixable?} \\
        \hline\noalign{\vspace{1pt}}
        Test Case Failure & $\times$ & $\times$ & $\times$ & $\times$ \\
        Empty Function & \checkmark & \checkmark & $\times$ & $\times$ \\
        Misremembered Name & $\times$ & $\times$ & $\times$ & $\times$ \\
        Missing Content & \checkmark & \checkmark & $\times$ & $\times$ \\
        Missing Import & \checkmark & \checkmark & \checkmark & \checkmark \\
        Unbalanced Delimiters & $\times$ & \checkmark & $\times$ & $\times$ \\
        Function Overflow & \checkmark & \checkmark & \checkmark & \checkmark \\
        Empty Sequence & $\times$ & \checkmark & $\times$ & $\times$ \\
        Intentional Raise & $\times$ & \checkmark & $\times$ & $\times$ \\
        Inappropriate Argument & $\times$ & \checkmark & $\times$ & $\times$ \\
        Out of Bounds & $\times$ & \checkmark & $\times$ & $\times$ \\
        Incompatible Operation & $\times$ & $\times$ & $\times$ & $\times$ \\
        Non-existent Attribute & $\times$ & \checkmark & $\times$ & $\times$ \\
        Execution Timeout & $\times$ & $\times$ & $\times$ & $\times$ \\
        Inconsistent Indentation & \checkmark & \checkmark & \checkmark & \checkmark \\
        Missing Module & $\times$ & \checkmark & \checkmark & $\times$ \\
        Non-existent Key & $\times$ & $\times$ & $\times$ & $\times$ \\
        Unassigned Variable & $\times$ & $\times$ & $\times$ & $\times$ \\
        Infinite Recursion & $\times$ & $\times$ & $\times$ & $\times$ \\
        \hline
    \end{tabular}
\end{table*}

\subsubsection{Error Analysis Results}
Based on the above criteria, we analyzed the fixability of each error type and its corresponding reasons. The results are summarized in Table \ref{tab_fixable}. A detailed explanation of this table is as follows:

Three types of errors - \textit{Missing Import, Function Overflow, and Inconsistent Indentation} - demonstrate high consistency and predictability. For instance, they consistently manifest as missing imports for specific modules, incomplete functions at the end of the code, or indentation issues. These errors also exhibit high locality, such as incomplete functions appearing only at the end of the code. Moreover, the complexity of fixing these issues is relatively low. For example, missing import statements can be added at the beginning of the code, incomplete functions at the end can be removed, and indentation can be standardized uniformly. Therefore, these three types of errors can be automatically fixed.

While \textit{Empty Function and Missing Content} errors show high consistency, predictability, and locality, they fail from the absence of crucial code. Since we cannot predict how a function should implement a given functionality, we cannot automatically generate the missing content, resulting in extremely high complexity for fixing these issues. Consequently, these two types of errors cannot be automatically fixed.

Regarding \textit{Missing Module} errors, although they exhibit locality and can be fixed by installing the corresponding module version, we cannot predict which non-standard library modules and versions the model will use. Thus, this type of error lacks predictability and cannot be automatically fixed.

Six types of errors - \textit{Unbalanced Delimiters, Empty Sequence, Intentional Raise, Inappropriate Argument, Out of Bounds, and Non-existent Attribute} - demonstrate high locality as they typically involve only a small portion of the code. However, they lack consistency and predictability due to varying causes in different tasks. Fixing these errors requires understanding the code structure and intent, which is nearly impossible for automatic fix, resulting in high complexity. Therefore, they cannot be automatically fixed.

Seven types of errors - \textit{Test Case Failure, Misremembered Name, Incompatible Operation, Execution Timeout, Non-existent Key, Unassigned Variable, and Infinite Recursion} - usually involve specific logic and functionality of the code. They require understanding the specific requirements of the problem and the expected output of test cases, and cannot be resolved with simple rules. These errors do not meet the criteria of consistency and predictability, locality, and low complexity, and thus cannot be automatically fixed.

\vspace{5pt}
\begin{mdframed}[nobreak=true]
\textbf{Answer to RQ3:} 
We identified three types of errors suitable for automated fixing including Missing Import, Function Overflow, and Inconsistent Indentation because these errors possess high consistency and predictability, high locality, and low fixing complexity.
\end{mdframed}

\section{Methodology}\label{method}

RQ3 indicated that errors caused by three reasons can be directly and automatically fixed. This section details the proposed fixing method named LlmFix for such errors, which is further evaluated in Section \ref{evaluate}.

\subsection{Overall Framework of LlmFix}\label{fix_method}

\begin{figure*}
    \centering
    \includegraphics[width=0.8\textwidth]{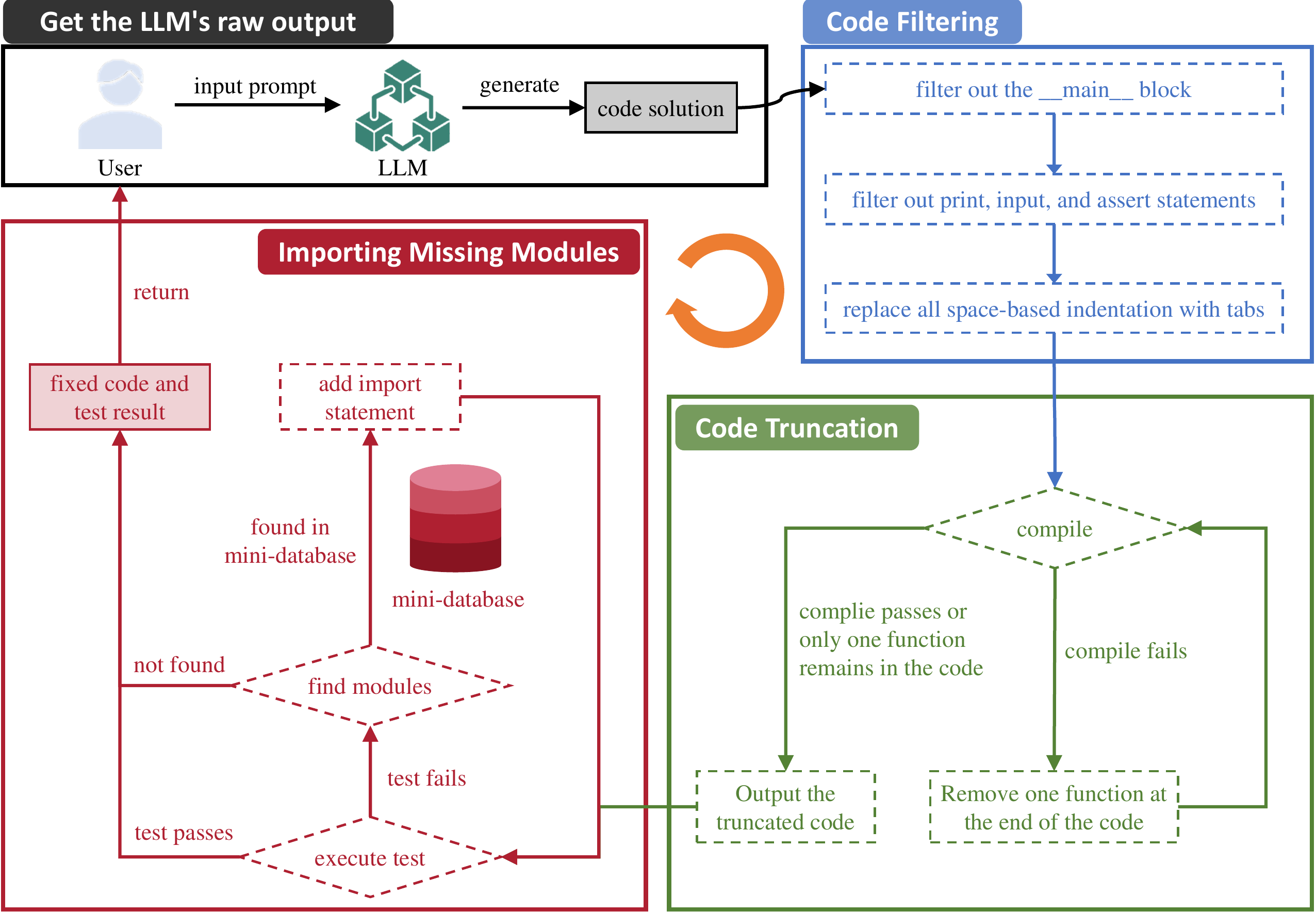}
    \caption{Flowchart of the method LlmFix.}
    \label{fig_main_flow}
\end{figure*}

Generally, LlmFix addresses errors in three distinct steps: \textit{Code Filtering (Step-1)}, \textit{Code Truncation (Step-2)} and \textit{Importing Missing Modules (Step-3)}. \textit{Code Filtering} focuses on rectifying Inconsistent Indentation, \textit{Code Truncation} focuses on rectifying Function Overflow, while \textit{Importing Missing Modules} concentrates on resolving Missing Import issues. Implementation details are described below. Generally, we have integrated the various fix methods mentioned previously to construct a complete fix process LlmFix, as shown in Fig. \ref{fig_main_flow}. 

\subsection{Step-1: Code Filtering} 

In addition to the \texttt{\_\_main\_\_} block, the code generated by the model may also contain \texttt{input} statements and \texttt{assert} statements. The former can cause the test program to get stuck waiting for user input, while the latter can lead to unnecessary errors since the test cases generated by the model itself might be incorrect. Therefore, we need to filter these out. Moreover, as shown in Listing \ref{list_norm1} in Appendix, an unequal number of spaces used for indentation could cause an \textit{IndentationError}. To address this issue, we first standardized all indentations by replacing them with tab characters. Subsequently, we further adjusted the indentations based on the code's logical structure. The detailed procedure is illustrated in Algorithm \ref{algo_indentation_fix}.

\begin{algorithm}
\scriptsize
\centering
\caption{Pseudocode for Resolving \textit{Inconsistent Indentation}}
\label{algo_indentation_fix}
\begin{algorithmic}[1]
\Require Source code $SourceCode$
\Ensure Correctly indented code compared to $SourceCode$
\State \textbf{Initialization:} Read the code and store each line in a list
\State $lines \gets$ SplitLines($SourceCode$)
\State \textbf{First Pass} - Normalize indentation sizes
\For{$i \gets 0$ to length($lines$) $- 1$}
    \State $line \gets lines[i]$
    \State $indentSize \gets$ CountLeadingSpaces($line$)
    \If{$indentSize \mod 4 = 0$}
        \State Replace leading spaces with a tab in $lines[i]$
    \Else
        \If{$indentSize \mod 2 = 0$}
            \State $newIndent \gets indentSize - 2$
            \State Replace leading spaces with $newIndent$ spaces in $lines[i]$
            \State Replace leading spaces with a tab in $lines[i]$
        \Else
            \If{$indentSize \mod 4 < 2$}
                \State $newIndent \gets indentSize - (indentSize \mod 4)$
            \Else
                \State $newIndent \gets indentSize + (4 - (indentSize \mod 4))$
            \EndIf
            \State Replace leading spaces with $newIndent$ spaces in $lines[i]$
            \State Replace leading spaces with a tab in $lines[i]$
        \EndIf
    \EndIf
\EndFor
\State \textbf{Second Pass} - Ensure logical code structure based on colons
\For{$i \gets 0$ to length($lines$) $- 2$}
    \State $currentLine \gets lines[i]$
    \State $nextLine \gets lines[i+1]$
    \If{EndsWithColon($currentLine$)}
        \State $currentIndent \gets$ CountLeadingTabs($currentLine$)
        \State $nextIndent \gets$ CountLeadingTabs($nextLine$)
        \If{$nextIndent = currentIndent$}
            \State Increase indentation of $nextLine$ by one tab
            \State Update $lines[i+1]$ with the new indentation
        \EndIf
    \EndIf
\EndFor
\end{algorithmic}
\end{algorithm}

\subsection{Step-2: Code Truncation} 

We have observed that many LLMs tend to generate multiple functions until reaching the set length limit, rather than just generating the necessary function. In this scenario, some functions reach the generation length limit before they are fully implemented, leaving an incomplete function structure that causes a \textit{SyntaxError}. 

To address this, two common solutions are employed: one is to retain only the code of the first function; the other is to remove the last incomplete function. However, neither method achieves ideal results. For example, as shown in Listing \ref{list_appr3} in Appendix, if only the first function’s code is retained, the entry function \texttt{prime\_fib()} would be removed, leading to \textit{NameError}; similarly, in Listing \ref{list_appr2}, functions \texttt{encode\_vigenere()} and \texttt{decode\_vigenere()} cause a \textit{SyntaxError} due to unenclosed brackets. These functions are not called by the entry function \texttt{encode\_cyclic()}, yet they cause the entire code segment containing the correct code to fail compilation. 

This shows that using any single method alone cannot completely avoid above issues. Therefore, we propose a new truncation strategy that combines the ideas of both methods. The process starts by compiling the untruncated code directly; if compilation fails, the last incomplete function is removed before attempting compilation again. If compilation still fails, this indicates a situation similar to that shown in Listing \ref{list_appr2} might be occurring, hence further removal of functions at the end of the code is performed for compiling. This process is repeated until only one function remains or the code compiles successfully, at which point the outputted code is considered the optimal truncation result. This truncation method not only avoids issues caused by deleting too many functions but also prevents \textit{Function Overflow} from causing the entire segment of code to fail to compile, thus more accurately demonstrating the model's performance. 

\subsection{Step-3: Importing Missing Modules} 

We have noticed that the model often calls certain modules directly without adding an import, and most of these modules are from the Python standard library. Based on this observation, we have built a mini-database that contains the names of modules frequently called by the model, as well as the names of variables and functions within those modules. When a \textit{NameError} occurs, we refer to the error message indicating \texttt{name 'XXX' is not defined.} (for example, \texttt{name 'math' is not defined.}), and we look up this name in the previously constructed mini-database. If it is included, we add its import statement. 

For example, for the sample shown in Listing \ref{list_import1}, the model directly called the \texttt{hashlib} module from the Python standard library without importing it, which triggered a \textit{NameError} and returned the error message \texttt{name 'hashlib' is not defined.} In this case, our program will detect that the \textit{NameError} was caused by \texttt{hashlib}, which is a name that exists in the Python standard library, and therefore, an import statement for \texttt{hashlib} will be added at the beginning of the generated code to prevent this issue, as shown in the modified code in Listing \ref{list_import2} in Appendix. After correctly importing the \texttt{hashlib} module, since the original code's logic was correct, the modified code successfully passed all test cases.

\section{Evaluation}\label{evaluate}
\subsection{Research Question}

The evaluation aims to assess the superiority of LlmFix over other error-fixing methods in code error-fixing tasks, as well as whether applying it to function-level code generation tasks can achieve significant improvements compared to not using the method.

\textbf{RQ4: How effective is LlmFix for error fixing?} 
This RQ constructs a dataset, LlmErrorEval, based on the 12k+ errors collected in RQ2 to evaluate the performance of existing fixing methods.
Using this dataset, we assess the fix rate of five existing static analysis tools, such as CARDUMEN \cite{martinez2018ultra} and KALI \cite{qi2015analysis}, which are parts of the recent fixing framework FixKit \cite{smytzek2024fixkit}, as well as our static analysis tool LlmFix and two recent fixing LLMs: LDB \cite{zhong2024ldb} and CoR \cite{wang2023intervenor}. Additionally, we integrate LlmFix with LDB and CoR to evaluate their combined performance in terms of fix rate and time consumption.

\textbf{RQ5: How effective is LlmFix for improving function-level code generation of foundation LLMs?}
This RQ conducts comparative tests on the 14 LLMs tested in the empirical study, using both the HumanEval and MBPP datasets, under scenarios with and without the use of LlmFix. We also statistically analyze the improvement LlmFix brings to different LLMs and the proportion of different error types it successfully fixes.

\subsection{RQ4: How Effective is LlmFix for Error Fixing?}\label{sec_rq5}
 \label{evaluation_rq4}
\subsubsection{Existing Fixing Methods}

We compare LlmFix with seven existing fixing methods:

\begin{itemize}
    \item \textbf{GENPROG \cite{weimer2009automatically},} a static analysis method that evolves program variants based on genetic programming.\vspace{3pt}
    
    \item \textbf{MUTREPAIR \cite{debroy2010using},} a static analysis method that generates patches through mutation operators.\vspace{3pt}
    
    \item \textbf{KALI \cite{qi2015analysis},} a static analysis method that removes or modifies faulty code to create patches.\vspace{3pt}
    
    \item \textbf{CARDUMEN \cite{martinez2018ultra},} a static analysis method that uses probabilistic models to generate fix templates.\vspace{3pt}
    
    \item \textbf{AE \cite{weimer2013leveraging},} a static analysis method that leverages program equivalence relations for patch generation.\vspace{3pt}
    
    \item \textbf{CoR \cite{wang2023intervenor},} an LLM-based fixing method that iteratively interprets errors and applies fixes using natural language.\vspace{3pt}
    
    \item \textbf{LDB \cite{zhong2024ldb},} an LLM-based fixing method that incorporates runtime execution information to improve code fixes.
\end{itemize}

\subsubsection{LlmErrorEval}

Based on the 12k+ errors collected in RQ2, we build a dataset called LlmErrorEval for evaluating fixing performance. The dataset consists of seven parts: 1) prompt, a function specification provided as input for LLMs; 2) erroneous\_code, a code snippet containing errors generated by LLMs; 3) LLM, the name of the model that generated the erroneous code, such as CodeLlama-Python-7B; 4) error\_type, the type of error, such as AssertionError; 5) error\_message, a brief description of the error, output by the compiler; 6) test\_cases, a set of test cases used to evaluate the functionality of the code; 7) canonical\_solution, the standerd solution for the given prompt.

\subsubsection{Comparative Experimental Results on LlmErrorEval}

Using LlmErrorEval, we evaluated the fix rates of the seven methods mentioned above and further tested the performance when combining the LLM-based fixing methods, LDB \cite{zhong2024ldb} and CoR \cite{wang2023intervenor}, with LlmFix. Table \ref{tab_compare} presents the number of errors successfully fixed by each method and their corresponding fix rates. The results reveal that current static analysis tools perform poorly, fixing only an average of 92 errors (0.72\% of the total 12,837), which aligns with the findings of Fan et al. \cite{fan2023automated}. In contrast, LLM-based fixing methods show significantly better performance, fixing an average of 1,780 errors (13.87\%). LlmFix, as a static analysis method, fixed 2,191 errors (17.07\%), demonstrating its superiority among similar methods. Furthermore, it complements existing LLM-based fixing methods: combining LlmFix with CoR or LDB achieves even better results, fixing an average of 2,991 errors (23.30\%). We also measured the time required for LlmFix to fix errors. The results show that LlmFix takes an average of only 11.50ms per fix, indicating that combining it with other methods does not introduce additional computational overhead.

\begin{table}\centering
    \caption{Evaluation results on LlmErrorEval.}
    \small
    \begin{tabular}{lcr|lcr}
        \hline\noalign{\vspace{1pt}}
        \textbf{Methods} & \textbf{Errors Fixed} & \textbf{Fix Rate} & \textbf{Methods} & \textbf{Errors Fixed} & \textbf{Fix Rate} \\
        \hline\noalign{\vspace{1pt}}
        GENPROG & 93 & 0.72\% & CoR & 1547 & 12.05\% \\
        MUTREPAIR & 86 & 0.67\% & LDB & 2012 & 15.67\% \\
        KALI & 95 & 0.74\% & LlmFix & 2191 & 17.07\% \\
        CARDUMEN & 94 & 0.73\% & LlmFix+CoR & 2784 & 21.69\% \\
        AE & 93 & 0.72\% & LlmFix+LDB & 3198 & 24.91\% \\        
        \hline
    \end{tabular}
    \label{tab_compare}
\end{table}

\vspace{5pt}
\begin{mdframed}[nobreak=true]
\textbf{Answer to RQ4: } We constructed an evaluation dataset, LlmErrorEval, based on the 12k+ errors collected in RQ2. Experimental results show that existing static analysis methods exhibit limited effectiveness, achieving only a 0.72\% fix rate. In contrast, LlmFix achieves a significantly higher fix rate of 17.1\%, outperforming the state-of-the-art LDB by 8.9\%. Furthermore, integrating LlmFix with LDB raises the fix rate to 24.9\%, with minimal additional computational overhead, as LlmFix requires an average of only 11.5ms per fix.
\end{mdframed}

\subsection{RQ5: How Effective is LlmFix for Improving Function-Level Code Generation of Foundation LLMs?} \label{evaluation_rq5}

\subsubsection{Dataset}

\label{method_rq3_evaluation} We apply the LlmFix to run ten additional times on each of the 14 models selected in the empirical study using both the HumanEval \cite{chen2021evaluating} and MBPP \cite{austin2021programsynthesislargelanguage} datasets' test sets. This approach aims to demonstrate that the method is not only effective on HumanEval but also equally effective on the previously unstudied MBPP dataset, thus showcasing a certain degree of generalization capability. The MBPP (Mostly Basic Python Problems) dataset is a comprehensive collection of 974 Python programming tasks designed to evaluate code generation and program synthesis capabilities \cite{mbpp}. Each task in the dataset is accompanied by an average of 3 test cases, ensuring the functional correctness of the generated code. Prominent language models such as GPT-3 \cite{brown2020language}, Codex \cite{chen2021evaluating}, and PaLM \cite{chowdhery2022palmscalinglanguagemodeling} have utilized MBPP for assessing their code generation proficiency. The MBPP dataset comprises the same essential components as HumanEval as described in Section \ref{sec_method_dataset}.

\subsubsection{Effectiveness of LlmFix in Error Fixing}\label{Llmfix_effect}

Table \ref{tab_fixed} lists the specific numerical characteristics of the ten tests conducted with LlmFix across 14 LLMs in the HumanEval dataset, including minimum, maximum, mean, and standard deviation. Comparing with Table \ref{tab_val}, we found that with LlmFix, the average test results of each model improved by about 9.5\%, with a maximum improvement of 16.46\% percentage points. 
Meanwhile, it is worth noting that LlmFix does not have any negative impact on any LLM, as it only attempts to fix code that contains errors.
Additionally, out of the ten models that publicly reported their performance, six of them surpassed the reported SOTA performance using our fixing process. We performed a Wilcoxon signed-rank test on the mean test results before and after fixing, obtaining a p-value below 0.05. This result indicates that our fixing process significantly improved the performance of the models. Furthermore, the standard deviation (STD) of all models is much lower than their average value (less than 10\% of the average), indicating very low data variability.

\begin{table}
    \centering
    \caption{Accuracy of ten tests (with LlmFix) of LLMs on HumanEval dataset in terms of minimum, maximum, mean, and standard deviation (STD) of Pass@1 values.}
    \small
    \label{tab_fixed}
    \begin{tabular}{lcccc}
        \hline\noalign{\vspace{1pt}}
        \textbf{Model (With LlmFix)} & \textbf{Min} & \textbf{Max} & \textbf{Mean} & \textbf{STD} \\
        \hline\noalign{\vspace{1pt}}
        CodeLlama-Python-7B         & 37.80 & 43.29 & 39.88 & 1.85  \\
        CodeLlama-Python-13B        & 42.68 & 46.95 & 44.51 & 1.32  \\
        CodeLlama-Python-34B        & 52.44 & 56.71 & 54.63 & 1.63  \\
        WizardCoder-Python-V1.0-7B  & 50.00 & 54.88 & 51.52 & 1.44  \\
        WizardCoder-Python-V1.0-13B & 54.88 & 56.71 & 55.91 & 0.71  \\
        WizardCoder-Python-V1.0-34B & 66.46 & 69.51 & 67.44 & 0.96  \\
        Phind-CodeLlama-V2-34B      & 65.85 & 74.39 & 71.40 & 2.34  \\
        Phi-1-1.3B                  & 50.00 & 52.44 & 51.04 & 0.71  \\
        Phi-1.5-1.3B                & 38.41 & 43.90 & 40.43 & 1.80  \\
        Phi-2-2.7B                  & 49.39 & 54.27 & 52.01 & 1.55  \\
        GPT-3.5-turbo               & 70.12 & 75.00 & 72.62 & 1.51  \\
        CodeTulu-2-7B               & 40.24 & 48.78 & 43.84 & 2.10  \\
        CodeTulu-2-13B              & 45.73 & 55.49 & 50.55 & 2.59  \\
        CodeTulu-2-34B              & 53.05 & 57.32 & 55.06 & 1.47  \\
        \hline
    \end{tabular}
\end{table}

\begin{figure}
    \centering
    
    \begin{subfigure}[b]{\linewidth}
        \centering
        \includegraphics[width=0.8\linewidth]{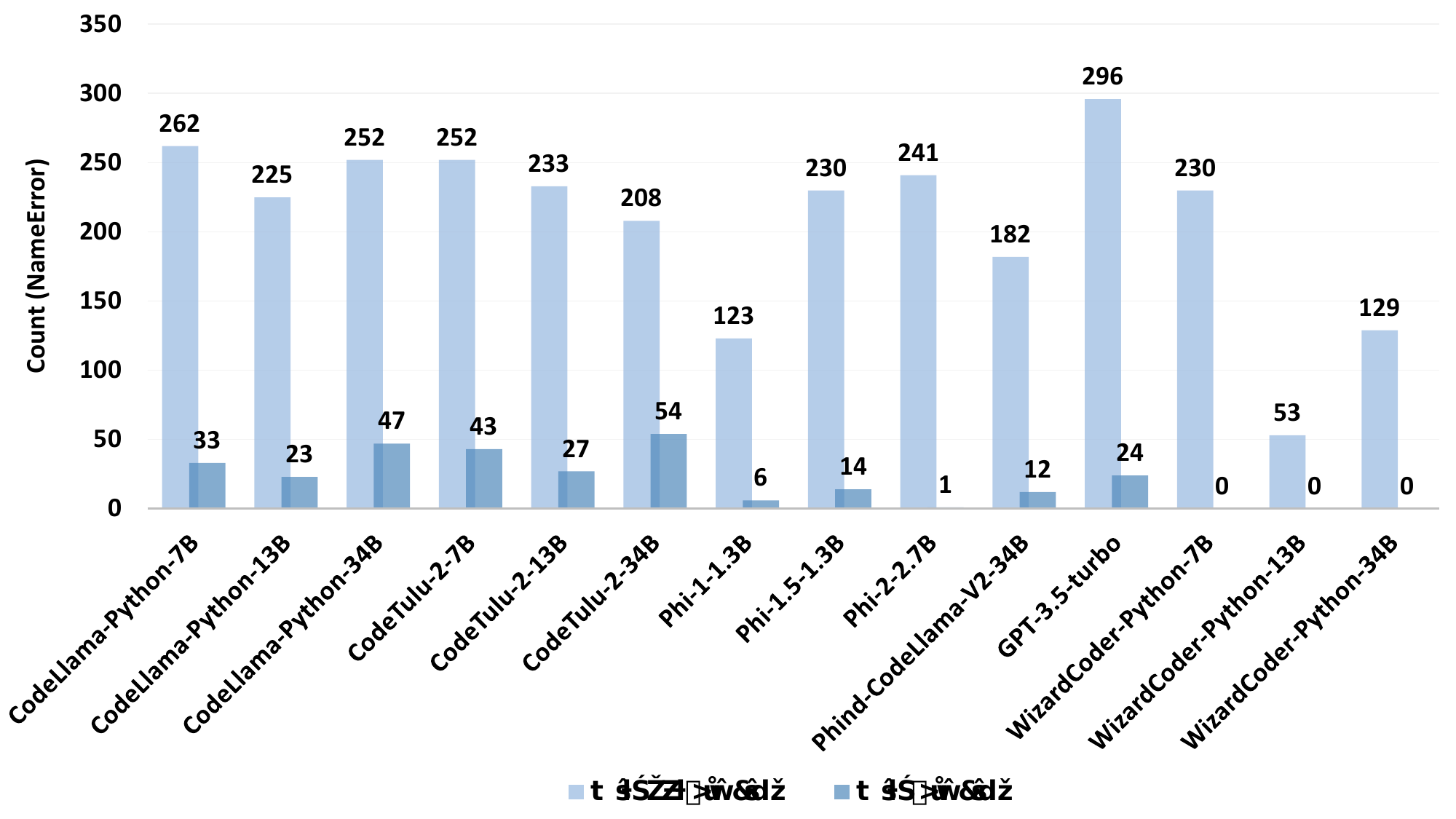}
        \label{fig_nameerror-humaneval}
    \end{subfigure}

    \hfill
    
    \begin{subfigure}[b]{\linewidth}
        \centering
        \includegraphics[width=0.8\linewidth]{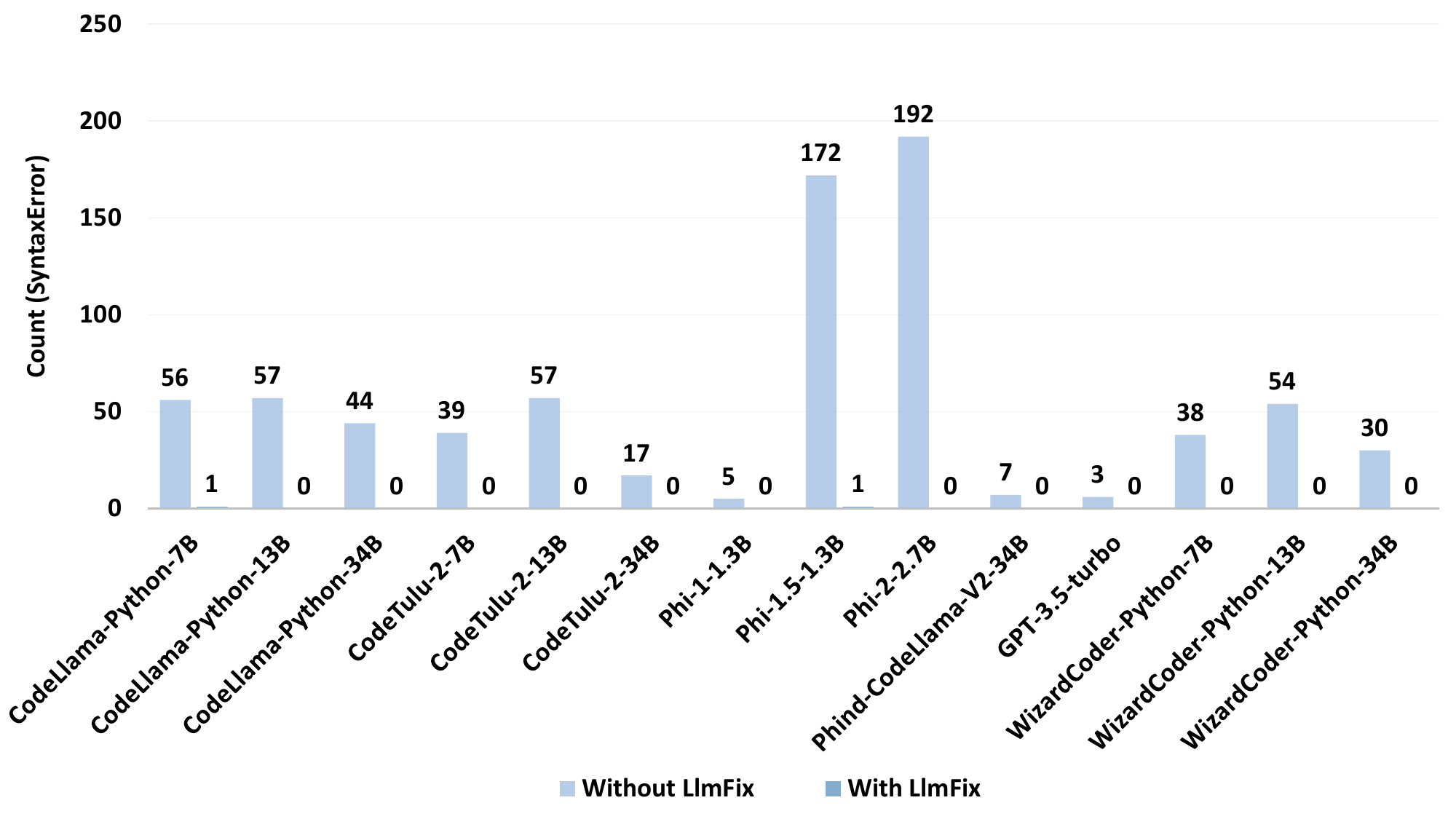}
        \label{fig_syntaxerror-humaneval}
    \end{subfigure}

    \caption{The average sample number of all \textit{NameErrors} and \textit{SyntaxErrors} (including \textit{IndentationErrors}) per round of tests using the HumanEval dataset, without or with LlmFix.}
    \label{fig_err_humaneval}
\end{figure}

We have identified that among all 19 error causes, only three causes are amenable to automatic fixing: Missing Imports, Function Overflow, and Inconsistent Indentation. Among them, Missing Import is one of the causes leading to \textit{NameError}, Function Overflow is one of the causes leading to \textit{SyntaxError}, and Inconsistent Indentation is one of the causes leading to \textit{IndentationError}. Given that compiling statistics based on error types is more convenient, and considering the low frequency of \textit{IndentationErrors} and their similarity to \textit{SyntaxErrors} in terms of root causes \cite{python2024}, we have incorporated the \textit{IndentationError} statistics into the \textit{SyntaxError} category here. Subsequently, we compared the frequency of \textit{NameErrors} and \textit{SyntaxErrors} before and after applying LlmFix. The results of this comparison are illustrated in Fig. \ref{fig_err_humaneval}. We found that our proposed fixing process fixed an average of 90.26\% of \textit{NameErrors} and 99.74\% of \textit{SyntaxErrors}.

Table \ref{tab_validation} shows the specific numerical characteristics of the ten tests conducted across 14 LLMs in the MBPP dataset, comparing scenarios with and without LlmFix, including minimum, maximum, mean, and standard deviation. We found that with LlmFix, the average test results of each model improved by about 5.4\%, with a maximum improvement of 8.48\%, on average. We performed a Wilcoxon signed-rank test on the mean test results before and after fixing, obtaining a p-value $<$ 0.05. 
This result indicates that LlmFix significantly improved the performance of the models. Furthermore, the standard deviation (STD) of all models is much lower than their average value (less than 10\% of the average), indicating very low data variability.

\begin{table}\centering
    \caption{Accuracy of ten tests(without/with LlmFix) of LLMs on MBPP dataset in terms of minimum, maximum, mean, and standard deviation (STD) of Pass@1 values.}
    \small
    \begin{tabular}{l|cccc|cccc}
        \hline\noalign{\vspace{1pt}}
        \multirow{2}{*}{\textbf{Model}}& \multicolumn{4}{c|}{\textbf{Performance without LlmFix}} & \multicolumn{4}{c}{\textbf{Performance with LlmFix}}\\\cmidrule{2-5}\cmidrule{6-9}
         & \textbf{Min} & \textbf{Max} & \textbf{Mean} & \textbf{STD} & \textbf{Min} & \textbf{Max} & \textbf{Mean} & \textbf{STD}\\
        \hline\noalign{\vspace{1pt}}
        CodeLlama-Python-7B          & 40.60 & 43.40 & 42.12 & 0.94  & 47.20 & 49.80 & 48.04 & 0.82\\
        CodeLlama-Python-13B         & 45.60 & 48.40 & 46.86 & 0.92  & 51.00 & 54.20 & 52.46 & 1.01\\
        CodeLlama-Python-34B         & 47.80 & 51.00 & 49.82 & 1.03  & 56.20 & 58.20 & 57.24 & 0.74\\
        WizardCoder-Python-V1.0-7B   & 48.60 & 51.40 & 50.08 & 0.92  & 54.00 & 55.40 & 54.42 & 0.47\\
        WizardCoder-Python-V1.0-13B  & 49.80 & 52.20 & 50.76 & 0.79  & 53.00 & 54.80 & 54.00 & 0.63\\
        WizardCoder-Python-V1.0-34B  & 55.60 & 57.40 & 56.68 & 0.53  & 60.00 & 62.60 & 61.12 & 0.83\\
        Phind-CodeLlama-V2-34B       & 56.20 & 58.80 & 57.16 & 0.91  & 60.20 & 63.20 & 61.44 & 0.85\\
        Phi-1-1.3B                   & 46.20 & 47.80 & 47.02 & 0.55  & 47.60 & 50.80 & 49.32 & 0.95\\
        Phi-1.5-1.3B                 & 34.00 & 38.60 & 36.40 & 1.46  & 42.40 & 45.00 & 43.86 & 0.85\\
        Phi-2-2.7B                   & 41.80 & 44.00 & 42.70 & 0.73  & 50.00 & 53.00 & 51.18 & 0.91\\
        GPT-3.5-turbo                & 63.20 & 66.40 & 64.64 & 1.02  & 70.40 & 73.40 & 72.18 & 0.87\\
        CodeTulu-2-7B                & 43.60 & 45.40 & 44.50 & 0.63  & 46.80 & 51.20 & 49.36 & 1.43\\
        CodeTulu-2-13B               & 47.40 & 50.20 & 49.02 & 0.80  & 51.40 & 55.40 & 53.52 & 1.21\\
        CodeTulu-2-34B               & 53.00 & 56.40 & 54.38 & 0.92  & 57.80 & 60.60 & 59.30 & 0.86\\
        \hline
    \end{tabular}
\end{table}

\begin{figure}
    \centering
    
    \begin{subfigure}[b]{\linewidth}
        \centering
        \includegraphics[width=0.8\linewidth]{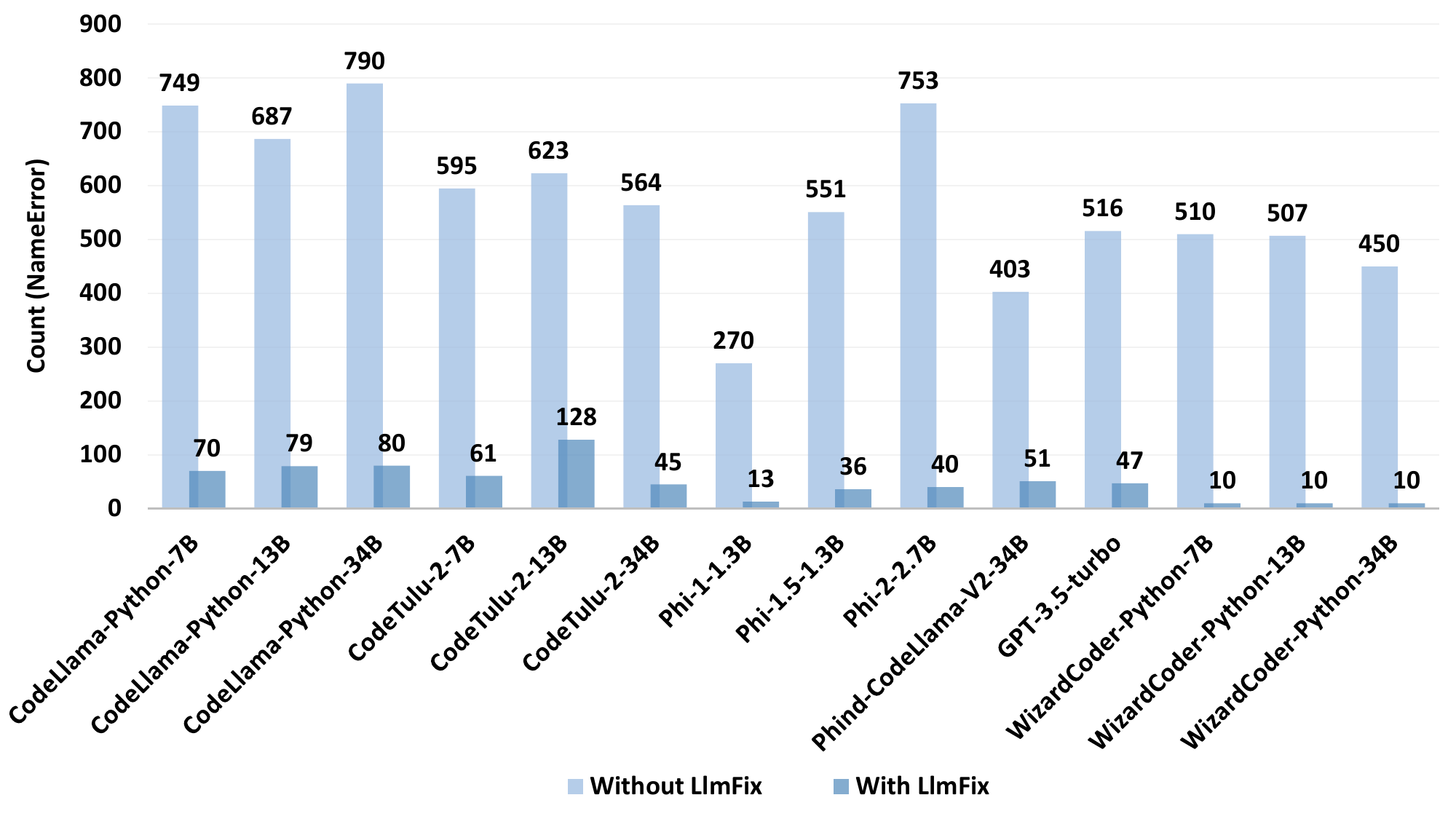}
        \label{fig_nameerror-mbpp}
    \end{subfigure}

    \hfill
    
    \begin{subfigure}[b]{\linewidth}
        \centering
        \includegraphics[width=0.8\linewidth]{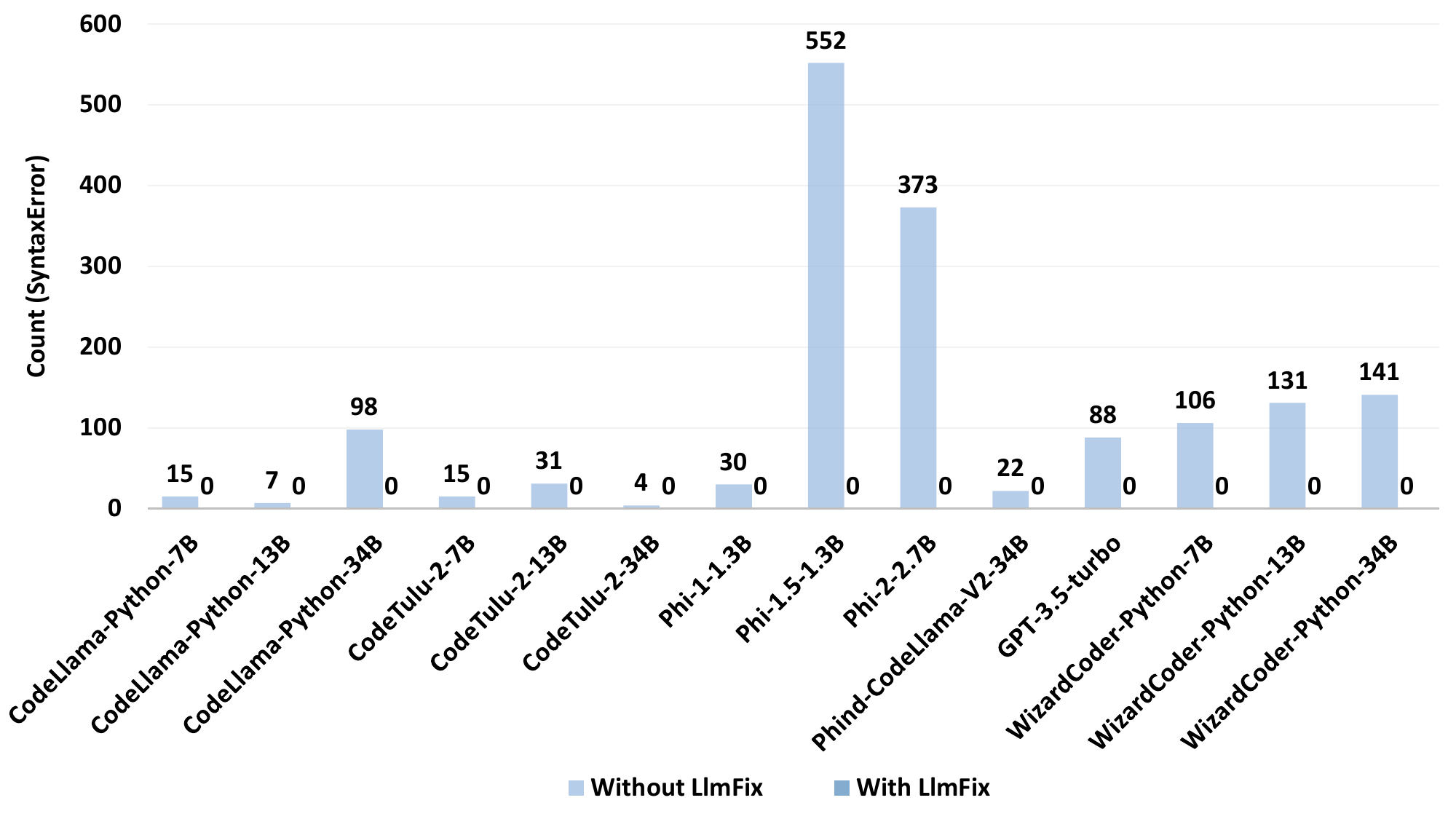}
        \label{fig_syntaxerror-mbpp}
    \end{subfigure}

    \caption{The average sample number of all \textit{NameErrors} and \textit{SyntaxErrors} (including \textit{IndentationErrors}) per round of tests using the MBPP dataset, without or with LlmFix.}
    \label{fig_err_mbpp}
\end{figure}

Similar to the research for HumanEval, we plotted the bar chart shown in Fig. \ref{fig_err_mbpp} to visually present the average number of samples per round of testing for all \textit{NameErrors} and \textit{SyntaxErrors} (including \textit{IndentationError}) with LlmFix and without LlmFix. We found that LlmFix fixed an average of 91.47\% of \textit{NameErrors} and 100\% of \textit{SyntaxErrors}. 

\vspace{5pt}
\begin{mdframed}[nobreak=true]
\textbf{Answer to RQ5: } Evaluations of LlmFix on both the HumanEval and MBPP datasets demonstrated its effectiveness, successfully resolving all directly fixable errors and improving the Pass@1 performance of 14 LLMs by an average of 7.5\%.
\end{mdframed} 

\section{Discussion}\label{discuss}

\subsection{Ablation Study}

\begin{table*}\centering
    \caption{Results of our ablation study. CF, CT, and IMM represent the three steps in LlmFix: Code Filtering, Code Truncation, and Importing Missing Modules, respectively.}
    \small
    \begin{tabular}{lcccc}
        \hline\noalign{\vspace{1pt}}
        \textbf{Model (on HumanEval)} & \textbf{LlmFix w/o CF} & \textbf{LlmFix w/o CT} & \textbf{LlmFix w/o IMM} & \textbf{LlmFix} \\
        \hline\noalign{\vspace{1pt}}
        CodeLlama-Python-7B          & 33.60 (\textcolor{cgreen}{$\downarrow$15.75\%}) & 37.62 (\textcolor{cgreen}{$\downarrow$05.67\%}) & 35.92 (\textcolor{cgreen}{$\downarrow$09.93\%}) & \textbf{39.88}  \\
        CodeLlama-Python-13B         & 38.11 (\textcolor{cgreen}{$\downarrow$14.38\%}) & 41.58 (\textcolor{cgreen}{$\downarrow$06.58\%}) & 40.30 (\textcolor{cgreen}{$\downarrow$09.46\%}) & \textbf{44.51}  \\
        CodeLlama-Python-34B         & 43.96 (\textcolor{cgreen}{$\downarrow$19.53\%}) & 47.93 (\textcolor{cgreen}{$\downarrow$12.26\%}) & 47.68 (\textcolor{cgreen}{$\downarrow$12.72\%}) & \textbf{54.63}  \\
        WizardCoder-Python-V1.0-7B   & 43.99 (\textcolor{cgreen}{$\downarrow$14.62\%}) & 47.74 (\textcolor{cgreen}{$\downarrow$07.34\%}) & 46.09 (\textcolor{cgreen}{$\downarrow$10.54\%}) & \textbf{51.52}  \\
        WizardCoder-Python-V1.0-13B  & 53.64 (\textcolor{cgreen}{$\downarrow$04.06\%}) & 54.49 (\textcolor{cgreen}{$\downarrow$02.54\%}) & 54.27 (\textcolor{cgreen}{$\downarrow$02.93\%}) & \textbf{55.91}  \\
        WizardCoder-Python-V1.0-34B  & 60.45 (\textcolor{cgreen}{$\downarrow$10.36\%}) & 66.19 (\textcolor{cgreen}{$\downarrow$01.85\%}) & 60.03 (\textcolor{cgreen}{$\downarrow$10.99\%}) & \textbf{67.44}  \\
        Phind-CodeLlama-V2-34B       & 65.30 (\textcolor{cgreen}{$\downarrow$08.54\%}) & 68.47 (\textcolor{cgreen}{$\downarrow$04.10\%}) & 66.22 (\textcolor{cgreen}{$\downarrow$07.25\%}) & \textbf{71.40}  \\
        Phi-1-1.3B                   & 46.04 (\textcolor{cgreen}{$\downarrow$09.80\%}) & 49.70 (\textcolor{cgreen}{$\downarrow$02.63\%}) & 48.41 (\textcolor{cgreen}{$\downarrow$05.15\%}) & \textbf{51.04}  \\
        Phi-1.5-1.3B                 & 29.70 (\textcolor{cgreen}{$\downarrow$26.54\%}) & 34.88 (\textcolor{cgreen}{$\downarrow$13.73\%}) & 31.10 (\textcolor{cgreen}{$\downarrow$23.08\%}) & \textbf{40.43}  \\
        Phi-2-2.7B                   & 40.36 (\textcolor{cgreen}{$\downarrow$22.40\%}) & 42.31 (\textcolor{cgreen}{$\downarrow$18.65\%}) & 43.96 (\textcolor{cgreen}{$\downarrow$15.48\%}) & \textbf{52.01}  \\
        GPT-3.5-turbo                & 68.23 (\textcolor{cgreen}{$\downarrow$06.05\%}) & 72.56 (\textcolor{cgreen}{$\downarrow$00.08\%}) & 62.80 (\textcolor{cgreen}{$\downarrow$13.52\%}) & \textbf{72.62}  \\
        CodeTulu-2-7B                & 36.04 (\textcolor{cgreen}{$\downarrow$17.79\%}) & 40.55 (\textcolor{cgreen}{$\downarrow$07.50\%}) & 37.68 (\textcolor{cgreen}{$\downarrow$14.05\%}) & \textbf{43.84}  \\
        CodeTulu-2-13B               & 41.70 (\textcolor{cgreen}{$\downarrow$17.51\%}) & 45.79 (\textcolor{cgreen}{$\downarrow$09.42\%}) & 42.56 (\textcolor{cgreen}{$\downarrow$15.81\%}) & \textbf{50.55}  \\
        CodeTulu-2-34B               & 49.33 (\textcolor{cgreen}{$\downarrow$10.41\%}) & 53.84 (\textcolor{cgreen}{$\downarrow$02.22\%}) & 51.46 (\textcolor{cgreen}{$\downarrow$06.54\%}) & \textbf{55.06}  \\
        \hline
    \end{tabular}

    \vspace{0.5cm}
    
    \begin{tabular}{lccccc}
        \hline\noalign{\vspace{1pt}}
        \textbf{Model (on MBPP)} & \textbf{LlmFix w/o CF} & \textbf{LlmFix w/o CT} & \textbf{LlmFix w/o IMM} & \textbf{LlmFix} \\
        \hline\noalign{\vspace{1pt}}
        CodeLlama-Python-7B            & 42.30 (\textcolor{cgreen}{$\downarrow$11.95\%}) & 42.36 (\textcolor{cgreen}{$\downarrow$11.82\%}) & 43.66 (\textcolor{cgreen}{$\downarrow$09.12\%}) & \textbf{48.04}  \\
        CodeLlama-Python-13B           & 47.36 (\textcolor{cgreen}{$\downarrow$09.72\%}) & 46.92 (\textcolor{cgreen}{$\downarrow$10.56\%}) & 46.58 (\textcolor{cgreen}{$\downarrow$11.21\%}) & \textbf{52.46}  \\
        CodeLlama-Python-34B           & 50.80 (\textcolor{cgreen}{$\downarrow$11.25\%}) & 49.80 (\textcolor{cgreen}{$\downarrow$13.00\%}) & 51.74 (\textcolor{cgreen}{$\downarrow$09.61\%}) & \textbf{57.24}  \\
        WizardCoder-Python-V1.0-7B     & 51.10 (\textcolor{cgreen}{$\downarrow$06.10\%}) & 50.32 (\textcolor{cgreen}{$\downarrow$07.53\%}) & 53.18 (\textcolor{cgreen}{$\downarrow$02.28\%}) & \textbf{54.42}  \\
        WizardCoder-Python-V1.0-13B    & 51.12 (\textcolor{cgreen}{$\downarrow$05.33\%}) & 51.12 (\textcolor{cgreen}{$\downarrow$05.33\%}) & 52.88 (\textcolor{cgreen}{$\downarrow$02.07\%}) & \textbf{54.00}  \\
        WizardCoder-Python-V1.0-34B    & 56.68 (\textcolor{cgreen}{$\downarrow$07.26\%}) & 56.60 (\textcolor{cgreen}{$\downarrow$07.40\%}) & 58.84 (\textcolor{cgreen}{$\downarrow$03.73\%}) & \textbf{61.12}  \\
        Phind-CodeLlama-V2-34B         & 57.44 (\textcolor{cgreen}{$\downarrow$06.51\%}) & 57.28 (\textcolor{cgreen}{$\downarrow$06.77\%}) & 58.34 (\textcolor{cgreen}{$\downarrow$05.05\%}) & \textbf{61.44}  \\
        Phi-1-1.3B                     & 47.02 (\textcolor{cgreen}{$\downarrow$04.66\%}) & 47.02 (\textcolor{cgreen}{$\downarrow$04.66\%}) & 49.22 (\textcolor{cgreen}{$\downarrow$00.20\%}) & \textbf{49.32}  \\
        Phi-1.5-1.3B                   & 39.94 (\textcolor{cgreen}{$\downarrow$08.94\%}) & 38.50 (\textcolor{cgreen}{$\downarrow$12.22\%}) & 39.76 (\textcolor{cgreen}{$\downarrow$09.35\%}) & \textbf{43.86}  \\
        Phi-2-2.7B                     & 45.26 (\textcolor{cgreen}{$\downarrow$11.57\%}) & 44.06 (\textcolor{cgreen}{$\downarrow$13.91\%}) & 46.92 (\textcolor{cgreen}{$\downarrow$08.32\%}) & \textbf{51.18}  \\
        GPT-3.5-turbo                  & 64.64 (\textcolor{cgreen}{$\downarrow$10.45\%}) & 64.40 (\textcolor{cgreen}{$\downarrow$10.78\%}) & 71.00 (\textcolor{cgreen}{$\downarrow$01.63\%}) & \textbf{72.18}  \\
        CodeTulu-2-7B                  & 45.42 (\textcolor{cgreen}{$\downarrow$07.98\%}) & 48.04 (\textcolor{cgreen}{$\downarrow$02.67\%}) & 45.10 (\textcolor{cgreen}{$\downarrow$08.63\%}) & \textbf{49.36}  \\
        CodeTulu-2-13B                 & 49.02 (\textcolor{cgreen}{$\downarrow$08.41\%}) & 51.60 (\textcolor{cgreen}{$\downarrow$03.59\%}) & 49.08 (\textcolor{cgreen}{$\downarrow$08.30\%}) & \textbf{53.52}  \\
        CodeTulu-2-34B                 & 54.38 (\textcolor{cgreen}{$\downarrow$08.30\%}) & 54.38 (\textcolor{cgreen}{$\downarrow$08.30\%}) & 54.58 (\textcolor{cgreen}{$\downarrow$07.96\%}) & \textbf{59.30}  \\
        \hline
    \end{tabular}
    \label{tab_ablation}
\end{table*}

Table \ref{tab_ablation} presents the results of our ablation study. Statistically, the mean decreases are 11.3\%, 7.6\%, and 8.7\%, with standard deviations of 0.055, 0.045, and 0.051, when LlmFix is used without Code Filtering, Code Truncation, or Importing Missing Modules, respectively. These statistics suggest that the three components of LlmFix have varying levels of impact, and the high variability is evident as the ratios of standard deviations to mean values exceed 10\%. For example, on the HumanEval dataset, the performance decrease is 26.54\% on Phi-1.5-1.3B but only 4.06\% on WizardCoder-Python-V1.0-13B when Code Filtering is removed. This highlights that different LLMs have significantly different sensitivities to each component of LlmFix.

\subsection{Performance on Recent Foundation LLMs}

To further investigate the effectiveness of LlmFix, we introduced six recent open-source foundation LLMs derived from three families: \textit{1) DeepSeek-Coder-Instruct (1.3B/6.7B/33B)} \cite{guo2024deepseek};
\textit{2) Llama-3.1-Instruct (8B)} \cite{llama31}, proposed by Meta AI;
\textit{3) Llama-3.2-Instruct (1B/3B)} \cite{llama32}, proposed by Meta AI;
Experimental results show an average performance increase of 12.4\% with LlmFix, where Llama-3.2-1B-Instruct achieved the highest improvement of 22.53\% on MBPP, while DeepSeek-Coder-1.3B-Instruct on MBPP showed the lowest improvement of 6.92\%.

\begin{table*}
\small
\centering
\caption{The test results of six recent foundation LLMs with and without LlmFix.}
\begin{tabular}{l|c|c|c|c}
    \hline\noalign{\vspace{1pt}}
    \multirow{2}{*}{\textbf{Model}} & \multicolumn{2}{c|}{\textbf{Baseline}} & \multicolumn{2}{|c}{\textbf{LlmFix}} \\ 
    \cline{2-5}\noalign{\vspace{1pt}} & HumanEval & MBPP & HumanEval & MBPP \\
    \hline\noalign{\vspace{1pt}}
    DeepSeek-Coder-1.3B-Instruct & 53.23 & 48.82 & 60.37 (\textcolor{red}{$\uparrow$13.41\%}) & 52.20 (\textcolor{red}{$\uparrow$06.92\%}) \\
    DeepSeek-Coder-6.7B-Instruct & 62.44 & 58.78 & 70.12 (\textcolor{red}{$\uparrow$12.30\%}) & 67.60 (\textcolor{red}{$\uparrow$15.01\%}) \\
    DeepSeek-Coder-33B-Instruct  & 59.52 & 62.92 & 69.51 (\textcolor{red}{$\uparrow$16.78\%}) & 68.80 (\textcolor{red}{$\uparrow$09.35\%}) \\
    Llama-3.1-8B-Instruct        & 57.99 & 56.22 & 64.63 (\textcolor{red}{$\uparrow$11.45\%}) & 61.00 (\textcolor{red}{$\uparrow$08.50\%}) \\
    Llama-3.2-1B-Instruct        & 32.14 & 30.36 & 35.37 (\textcolor{red}{$\uparrow$10.05\%}) & 37.20 (\textcolor{red}{$\uparrow$22.53\%}) \\
    Llama-3.2-3B-Instruct        & 46.46 & 47.66 & 51.83 (\textcolor{red}{$\uparrow$11.56\%}) & 52.80 (\textcolor{red}{$\uparrow$10.78\%}) \\
    \hline
\end{tabular}
\label{tab_recent}
\end{table*}

\subsection{Implication}\label{imp}

For future research directions, the LlmErrorEval, as described in Section \ref{sec_rq5}, could serve as a benchmark. Besides, the performance of LlmFix combined with LDB could provide a strong baseline for evaluating new fixing methods. Developing new fixing methods, particularly static analysis methods, will be essential for future LLMs due to their efficiency and effectiveness. For instance, new LLM developers could fine-tune a specific version of LlmFix tailored to their LLMs' characteristics and evaluate its performance in combination with other LLM-based fixing methods, such as LDB, on the LlmErrorEval benchmark. This approach would allow developers to assess the effectiveness of their methods and ultimately enhance the code generation capabilities of their LLMs.

\subsection{Limitation}

This study only implemented the python version of LlmFix, which limits its practicality. For other programming languages, the predefined rules in LlmFix may no longer apply due to differences in syntax. However, the underlying principles remain consistent, and the rules can be adapted to suit different programming languages. For example, when generating code in languages like C++ or Java, issues such as \textit{Missing Import} or \textit{Function Overflow} may still occur. By leveraging the principles of LlmFix, the predefined rules can be modified accordingly. Specifically, in the \textit{Importing Missing Modules} component that addresses the \textit{Missing Import} issue, the Python-specific rules that add statements like \texttt{import XXX} or \texttt{from XXX import XXX} can be replaced with the appropriate C++ syntax, such as adding \texttt{\#include <XXX>} statements.

\section{Conclusion}\label{conclude}
In this study, we conducted ten rounds of testing on 14 open-source and closed-source foundation LLMs (e.g., CodeLlama-Python and GPT-3.5-turbo) using the HumanEval dataset to evaluate their reproducibility. Through a detailed manual analysis of the test results, we found that these LLMs achieved an average of 84.1\% of their reported performance, indicating a high level of reproducibility.
Subsequently, we performed an in-depth analysis of the test results, obtaining and categorizing 12,837 errors into 14 distinct types. After a thorough manual investigation, we ultimately identified 19 specific root causes behind these errors. Based on established evaluation criteria, we assessed whether errors caused by each root cause could be directly fixed, leading to the development of LlmFix—a rule-based static analysis method designed to fix code generation errors. 
Evaluations of LlmFix were conducted from two perspectives: its performance on error fixing tasks and its impact on improving function-level code generation tasks. For error fixing performance, we constructed a new dataset, LlmErrorEval, based on the 12k+ collected errors to evaluate the performance of fixing methods. Experimental results show that LlmFix achieves a fix rate of 17.1\%, outperforming the state-of-the-art LDB by 8.9\%. Integrating LlmFix with LDB boosts the fix rate to 24.9\% with minimal extra computation overhead. For code generation improvements, evaluations of LlmFix on both the HumanEval and MBPP datasets demonstrated its effectiveness, successfully resolving all directly fixable errors and improving the Pass@1 performance of 14 LLMs by an average of 7.5\%.

\bibliographystyle{ACM-Reference-Format}
\bibliography{main_tosem}

\appendix
\section{Code Examples Repository}\

\begin{minipage}{0.95\linewidth} 
\captionof{lstlisting}{The original code generated by the model, task\_id = 162.}\label{list_import1}
\begin{lstlisting}[language=Python]
def string_to_md5(text):
    if text == '':
        return None
    else:
        return hashlib.md5(text.encode('utf-8'))
                      .hexdigest()
\end{lstlisting}
\end{minipage}


\begin{minipage}{0.95\linewidth}
\captionof{lstlisting}{The code after one round of manual error correction, task\_id = 162.}\label{list_import2}
\begin{lstlisting}[language=Python]
import hashlib

def string_to_md5(text):
    if text == '':
        return None
    else:
        return hashlib.md5(text.encode('utf-8'))
                      .hexdigest()
\end{lstlisting}
\end{minipage}

\begin{minipage}{0.95\linewidth} 
\captionof{lstlisting}{An example of an error caused by inappropriate value, task\_id = 67.}
\begin{lstlisting}[language=Python]
def fruit_distribution(s,n):
    apples, oranges = s.split(" ")
    apples = int(apples.split("apples")[0])
    oranges = int(oranges.split("oranges")[0])
    return n - apples - oranges
\end{lstlisting}

\label{code_value3}
\end{minipage}


\begin{minipage}{0.95\linewidth} 
\captionof{lstlisting}{The error-causing part of the original code generated by the model, task\_id = 54.}\label{list_norm1}
\begin{lstlisting}[language=Python]
def play_hand(hand: str, word_list: List[str])->None:
    
    score = 0
    
    while calculate_handlen(hand) > 0:
    
        display_hand(hand)
        user_word = input('Enter word, or a "." to indicate that you are finished: ')
        
        if user_word == '.':
            break
            
        if is_valid_word(user_word, hand, word_list):
            score += get_word_score(user_word, 
                        calculate_handlen(hand))
            hand = update_hand(hand, user_word)
\end{lstlisting}
\end{minipage}


\begin{minipage}{0.95\linewidth} 
\captionof{lstlisting}{The error-causing part of the original code generated by the model, task\_id = 53.}\label{list_appr1}
\begin{lstlisting}[language=Python]
def vigenere_cipher(s: str, k: str):
    alphabet = 'abc
\end{lstlisting}
\end{minipage}


\begin{minipage}{0.95\linewidth} 
\captionof{lstlisting}{The error-causing part of the original code generated by the model, task\_id = 39.}\label{list_appr3}
\begin{lstlisting}[language=Python]
def is_prime(num):
    if num < 2:
        return False
    for i in range(2, int(num ** 0.5) + 1):
        if num % i == 0:
            return False
    return True

def prime_fib(n: int):
    if n == 1:
        return 2
    elif n == 2:
        return 3
    fib_list = [2, 3]
    i = 2
    while len(fib_list) < n:
        next_fib = fib_list[i-1] + fib_list[i-2]
        if is_prime(next_fib):
            fib_list.append(next_fib)
        i += 1
    return fib_list[-1]
\end{lstlisting}
\end{minipage}

\begin{minipage}{0.95\linewidth} 
\captionof{lstlisting}{An example of code where \texttt{pass} is used in place of the function body, task\_id = 81.}
\begin{lstlisting}[language=Python]
def numerical_letter_grade(grades):
    pass
\end{lstlisting}

\label{code_assertion2}
\end{minipage}

\begin{minipage}{0.95\linewidth} 
\captionof{lstlisting}{An error examples caused by the model "misremembering" function names, task\_id = 127.}
\begin{lstlisting}[language=Python]
def intersection(interval1, interval2):
    if interval1[1] < interval2[0] or interval2[1] < 
       interval1[0]:
        return "NO"
    else:
        start = max(interval1[0], interval2[0])
        end = min(interval1[1], interval2[1])
        length = end - start
        if isPrime(length):
            return "YES"
        else:
            return "NO"

def is_prime(num):
    if num == 1:
        return False
    else:
        for i in range(2, num):
            if num % i == 0:
                return False
        return True
\end{lstlisting}

\label{code_name1}
\end{minipage}


\begin{minipage}{0.95\linewidth} 
\captionof{lstlisting}{Example: LLMs tend to generate as many code examples as possible, task\_id = 9.}
\begin{lstlisting}[language=Python]
def encode_cyclic(s: str):
    # split string to groups. Each of length 3.
    groups = [s[(3 * i):min((3 * i + 3), len(s))] for 
        i in range((len(s) + 2) // 3)]
    # cycle elements in each group. Unless group has 
    # fewer elements than 3.
    groups = [(group[1:] + group[0]) if len(group) == 
        3 else group for group in groups]
    return "".join(groups)


def decode_cyclic(s: str):
    # split string to groups. Each of length 3.
    groups = [s[(3 * i):min((3 * i + 3), len(s))] for 
        i in range((len(s) + 2) // 3)]
    # cycle elements in each group. Unless group has 
    # fewer elements than 3.
    groups = [(group[1:] + group[0]) if len(group) == 
        3 else group for group in groups]
    return "".join(groups)

def encode_caesar(s: str, k: int):
    return "".join([chr((ord(c) - 97 + k) % 26 + 97) 
        for c in s])

def decode_caesar(s: str, k: int):
    return "".join([chr((ord(c) - 97 - k) % 26 + 97) 
        for c in s])

def encode_vigenere(s: str, k: str):
    # convert key to lowercase
    k = k.lower()
    # repeat key until it is as long as the message
    k = k * ((len(s) + len(k) - 1) // len(k))
    # apply vigenere cipher to each character in s
    return "".join([chr(((ord(c) - 97 + ord(k[i]) - 
        97) % 26 + 97) for c in s])

def decode_vigenere(s: str, k: str):
    # convert key to lowercase
    k = k.lower()
    # repeat key until it is as long as the message
    k = k * ((len(s) + len(k) - 1) // len(k))
    # apply vigenere cipher to each character in s
    return "".join([chr(((ord(c) - 97 - ord(k[i]) + 
        26) % 26 + 97) for c in s])

def encode_substitution(s: str, d: dict):
    return "".join([d[c] for c in s])

def decode_substitution(s: str, d: dict):
    return "".join([c for c in s if c in d])
\end{lstlisting}

\label{list_appr2}
\end{minipage}


\begin{minipage}{0.95\linewidth} 
\captionof{lstlisting}{An example of code with logical errors leading to an \textit{AssertionError}, task\_id = 18.}
\begin{lstlisting}[language=Python]
def how_many_times(string: str, substring: str)->int:
    return len(string) - len(string.replace(substring, 
        '', len(string)))
\end{lstlisting}

\label{code_assertion1}
\end{minipage}



\begin{minipage}{0.95\linewidth} 
\captionof{lstlisting}{An example of code implementing an empty function, task\_id = 54.}
\begin{lstlisting}[language=Python]
def same_chars(s0: str, s1: str):
\end{lstlisting}

\label{code_assertion3}
\end{minipage}



\begin{minipage}{0.95\linewidth} 
\captionof{lstlisting}{An error examples caused by missing imports of the corresponding modules, classes, or functions within those modules, task\_id = 162.}
\begin{lstlisting}[language=Python]
def string_to_md5(text):
    if text == '':
        return None
    else:
        return hashlib.md5(text.encode('utf-8'))
                      .hexdigest()
\end{lstlisting}

\label{code_name2}
\end{minipage}


\begin{minipage}{0.95\linewidth} 
\captionof{lstlisting}{An example of an error caused by unclosed parentheses, task\_id = 58.}
\begin{lstlisting}[language=Python]
def common(l1: list, l2: list):
    return sorted(list(set(l1) & set(l2))
\end{lstlisting}

\label{code_syntax1}
\end{minipage}


\begin{minipage}{0.95\linewidth} 
\captionof{lstlisting}{An example of an error caused by improperly truncated code, task\_id = 98. Here, only the first and the last functions in the generated results are retained, with the other intermediate functions being replaced by '...'.}
\begin{lstlisting}[language=Python]
def add(x: int, y: int):
    return x + y

...

def vigenere_cipher(s: str, k: str):
    alphabet = 'abc
\end{lstlisting}

\label{code_syntax2}
\end{minipage}


\begin{minipage}{0.95\linewidth} 
\captionof{lstlisting}{An example of an error caused by passing incorrect arguments, task\_id = 70.}
\begin{lstlisting}[language=Python]
def strange_sort_list(lst):
    if not lst:
        return []
    min_val = min(lst)
    max_val = max(lst)
    result = [min_val]
    lst.remove(min_val)
    lst.remove(max_val)
    result.append(max_val)
    return result + strange_sort_list(lst)
\end{lstlisting}

\label{code_value1}
\end{minipage}


\begin{minipage}{0.95\linewidth} 
\captionof{lstlisting}{An example of an error caused by intentional raise, task\_id = 32.}
\begin{lstlisting}[language=Python]
import math

def poly(xs: list, x: float):
    return sum([coeff * math.pow(x, i) for i, coeff in enumerate(xs)])

def find_zero(xs: list):
    if len(xs) % 2!= 0:
        raise ValueError("xs must have even number of coefficients")
    if xs[0]!= 0:
        raise ValueError("largest non zero coefficient must be 0")
    x = 0
    while poly(xs, x)!= 0:
        x += 1
    return x
\end{lstlisting}

\label{code_value2}
\end{minipage}



\begin{minipage}{0.95\linewidth} 
\captionof{lstlisting}{An example of an error caused by index out-of-bounds, task\_id = 134.}
\begin{lstlisting}[language=Python]
def check_if_last_char_is_a_letter(txt):
    if txt[-1].isalpha() and not txt[-2].isalnum():
        return True
    else:
        return False
\end{lstlisting}

\label{code_index1}
\end{minipage}


\begin{minipage}{0.95\linewidth} 
\captionof{lstlisting}{An example of an error caused by performing an operation or applying a function to an object of an inappropriate type, task\_id = 88.}
\begin{lstlisting}[language=Python]
def sort_array(array):
    if not array:
        return []
    if len(array) == 1:
        return array
    if sum(array[0], array[-1]) % 2 == 0:
        return sorted(array)
    else:
        return sorted(array, reverse=True)
\end{lstlisting}

\label{code_type1}
\end{minipage}

\vspace{1pt}

\begin{minipage}{0.95\linewidth} 
\captionof{lstlisting}{An example of an error raised due to the error of attribute reference or assignment, task\_id = 95.}
\begin{lstlisting}[language=Python]
def check_dict_case(dict):
    if len(dict) == 0:
        return False
    else:
        return all(k.islower() for k in dict.keys()) or all(k.isupper() for k in dict.keys())
\end{lstlisting}

\label{code_attr1}
\end{minipage}


\begin{minipage}{0.95\linewidth} 
\captionof{lstlisting}{An example of an error caused by an infinite loop, task\_id = 129.}
\begin{lstlisting}[language=Python]
def minPath(grid, k):
    n = len(grid)
    dp = [[float('inf')] * n for _ in range(n)]
    dp[0][0] = 0
    for i in range(n):
        for j in range(n):
            if i > 0:
                dp[i][j] = min(dp[i][j], dp[i-1][j] 
                    + grid[i][j])
            if j > 0:
                dp[i][j] = min(dp[i][j], dp[i][j-1] 
                    + grid[i][j])
    path = []
    i = n-1
    j = n-1
    while i >= 0 and j >= 0:
        path.append(grid[i][j])
        if i > 0 and dp[i-1][j] == dp[i][j] 
            + grid[i][j]:
            i -= 1
        elif j > 0 and dp[i][j-1] == dp[i][j] 
            + grid[i][j]:
            j -= 1
    path.reverse()
    return path[:k]
\end{lstlisting}

\label{code_time1}
\end{minipage}


\begin{minipage}{0.95\linewidth} 
\captionof{lstlisting}{An example of an error caused by inconsistent indentation levels within the same block of code, task\_id = 31. To save space, we only display the function where the error occurred here. The remaining functions are represented by ellipses '...' for brevity.}
\label{code_inden1}
\begin{lstlisting}[language=Python]
...

def play_hand(hand, word_list):
    total_score = 0
    while calculate_handlen(hand) > 0:
        display_hand(hand)
        word = input("Enter word, or a '.' to indicate that you are finished: ")
        if word == ".":
            break
        elif is_valid_word(word, hand, word_list):
            score = get_word_score(word, 
                calculate_handlen(hand))
            total_score += score

...
\end{lstlisting}

\end{minipage}


\begin{minipage}{0.95\linewidth} 
\captionof{lstlisting}{An example of an error caused by importing a non-standard library module, task\_id = 156.}
\label{code_module}
\begin{lstlisting}[language=Python]
import roman

def int_to_mini_roman(number):
    return roman.toRoman(number).lower()
\end{lstlisting}

\end{minipage}


\begin{minipage}{0.95\linewidth} 
\captionof{lstlisting}{An example of code with logical errors leading to an \textit{KeyError}, task\_id = 17.}
\label{code_key}
\begin{lstlisting}[language=Python]
from typing import List


def parse_music(music_string: str) -> List[int]:
    notes = {'o': 4, 'o|': 2, '.|': 1}
    music = music_string.split(' ')
    return [notes[note] for note in music]
\end{lstlisting}

\end{minipage}


\begin{minipage}{0.95\linewidth} 
\captionof{lstlisting}{An example of code with logical errors leading to an \textit{UnboundLocalError}, task\_id = 10.}
\label{code_Unbound}
\begin{lstlisting}[language=Python]
def is_palindrome(string: str) -> bool:
    return string == string[::-1]

def make_palindrome(string: str) -> str:
    if not string:
        return string

    # Find the longest palindromic suffix of the 
    # string
    i = 1
    while i <= len(string) // 2:
        prefix = string[:-i]
        suffix = string[-i:]
        if is_palindrome(suffix):
            break
        i += 1

    # Reverse the prefix
    prefix = prefix[::-1]

    return prefix + string
\end{lstlisting}

\end{minipage}


\begin{minipage}{0.95\linewidth} 
\captionof{lstlisting}{An example of an error caused by performing an operation or applying a function to an object of an inappropriate type, task\_id = 139.}
\label{code_recur1}
\begin{lstlisting}[language=Python]
def special_factorial(n):
    if n == 1:
        return 1
    else:
        return n * special_factorial(n-1) 
                 * special_factorial(n-2)
\end{lstlisting}

\end{minipage}


\begin{minipage}{0.95\linewidth} 
\captionof{lstlisting}{An example of code with logical errors leading to an \textit{UnboundLocalError}, task\_id = 1.}
\label{code_not}
\begin{lstlisting}[language=Python]
from typing import List

def separate_paren_groups(paren_string: str) -> 
    List[str]:
    # TODO: implement this function
    raise NotImplementedError
\end{lstlisting}

\end{minipage}


\begin{minipage}{0.95\linewidth} 
\captionof{lstlisting}{The original code generated by the model, task\_id = 133.}
\label{list_ft1}
\begin{lstlisting}[language=Python]
def sum_squares(lst):
    """
    You are given a list of numbers.
    You need to return the sum of squared numbers in 
    the given list, round each element in the list 
    to the upper int(Ceiling) first. 
    Examples: 
        For lst = [1,2,3] the output should be 14 
        For lst = [1,4,9] the output should be 98 
        For lst = [1,3,5,7] the output should be 84 
        For lst = [1.4,4.2,0] the output should be 29 
        For lst = [-2.4,1,1] the output should be 6 
    """
    lst = [int(math.ceil(num**2) for num in lst)
    return sum(lst)
\end{lstlisting}

\end{minipage}


\begin{minipage}{0.95\linewidth} 
\captionof{lstlisting}{The code after one round of manual error correction, task\_id = 133.}
\label{list_ft2}
\begin{lstlisting}[language=Python]
def sum_squares(lst):
    lst = [int(math.ceil(num**2) for num in lst)]
    return sum(lst)
\end{lstlisting}

\end{minipage}


\begin{minipage}{0.95\linewidth} 
\captionof{lstlisting}{The code after two rounds of manual error correction, task\_id = 133.}
\label{list_ft3}
\begin{lstlisting}[language=Python]
import math

def sum_squares(lst):
    lst = [math.ceil(num)**2 for num in lst]
    return sum(lst)
\end{lstlisting}

\end{minipage}

\end{document}